\documentclass[reqno, 11pt]{amsart}
\usepackage{mathrsfs}
\usepackage[round]{natbib}
\usepackage{amsmath}
\usepackage{amsthm}
\usepackage{amssymb, nicefrac}
\usepackage{graphicx, xcolor}
\usepackage{setspace}
\usepackage{mathtools}
\usepackage{parskip}
\usepackage{enumitem}

\calclayout

\providecommand{\tabularnewline}{\\}
\theoremstyle{plain}
\newtheorem{thm}{\protect\theoremname}
\providecommand{\theoremname}{Theorem}
\newtheorem{lem}{Lemma}
\newtheorem{defn}{Definition}
\begin{document}
\title[]{Copula estimation for nonsynchronous financial data}
\author[]{A\lowercase{rnab} C\lowercase{hakrabarti}$^{1}$ , R\lowercase{ituparna} S\lowercase{en}$^{2, \dag}$\\
$^{1}$M\lowercase{isra} C\lowercase{entre} \lowercase{for} F\lowercase{inancial} M\lowercase{arket and} E\lowercase{conomy},\\
I\lowercase{ndian} I\lowercase{nstitute of} M\lowercase{anagement} A\lowercase{hmedabad}, G\lowercase{ujarat}, I\lowercase{ndia}\\
$^{2}$I\lowercase{ndian} S\lowercase{tatistical} I\lowercase{nstitute}, B\lowercase{angalore}, I\lowercase{ndia}}
\thanks{\dag{\it Corresponding author}, {E-mail: \tt rsen@isibang.ac.in}, {Address: Indian Statistical Institute, Bangalore, 8th Mile, Mysore Road, RVCE Post, Bengaluru, Karnataka 560059}}
\maketitle
\begin{abstract}
Copula is a powerful tool to model multivariate data.  We  propose the modelling of intraday financial returns of multiple assets through copula. The problem originates due to the asynchronous nature of intraday financial data.  We propose a consistent estimator of the correlation coefficient in case of Elliptical copula and show that the plug-in copula estimator is uniformly convergent. For non-elliptical copulas, we capture the dependence through Kendall's Tau.  We demonstrate underestimation of the copula parameter and use a quadratic model to propose an improved estimator. In simulations, the proposed estimator reduces the bias significantly for a general class of copulas. We apply the proposed methods to real data of several stock prices.
\end{abstract}


\newpage

\section{Introduction}\label{sec:Introduction}

\begin{center}
\begin{figure}
\includegraphics[scale=0.4]{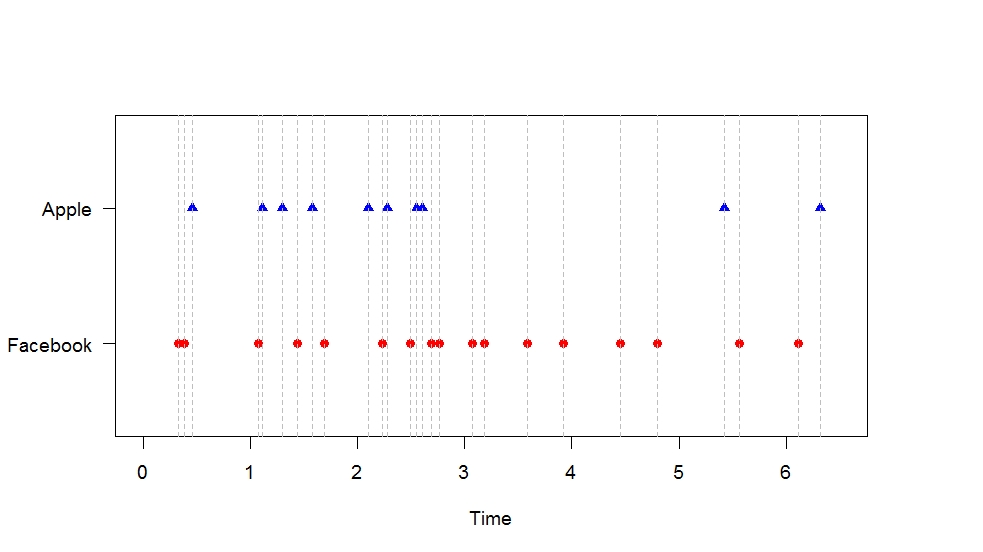}
\caption{Transaction times for two stocks Facebook and Apple for first few ticks on a particular day (10/05/2017). The scale on the x-axis is seconds.}\label{Figure 1}
\end{figure}
\end{center}

A very rich collection of market risk models have been developed and thoroughly investigated for intraday financial data. Although univariate modeling is important for addressing certain kinds of problems, it is not enough to unveil the nature and dynamics of the financial market. Interactions between different financial instruments are left out of univariate studies. If the companies belong to related business sectors or are owed by the same business house then such interactions can arise. High dependence between several constituent stocks of a portfolio can increase the probability of a large loss. So accurate estimation of the dependence between assets is of paramount importance. Correlation dynamics models, therefore, have become an important aspect of the theory and practice in finance. \emph{Correlation trading}, which is a trading activity to exploit the changes in dependence structure of financial assets, and \emph{correlation risk} that capture the exposure to losses due to changes in correlation, have attracted the attention of many practitioners, see \citet{krishnan2009correlation}. Accurate modelling of dependence is also important in a range of practical scenarios. For example, basket options are widely used although their accurate pricing is challenging. The primary reason is that they are cheaper to use for portfolio insurance. The cost-saving relies on the dependence structure between the assets, see \citet{salmon2006pricing}. In the actuarial world, as shown in \citet{embrechts2002correlation}, some Monte Carlo-based approach to joint modelling of risks, like Dynamic Financial Analysis, depends heavily on the dependence structure.
\citet{frey2002var} and \citet{breymann2003dependence} showed that the choice of model and correlation have a significant impact on the tail of the loss distribution and measures of extreme risks. It follows from the above discussion that we need accurate multivariate modeling and analysis. In order to perform multivariate analysis, we need multivariate data. This means that we need to have data for all $p (\geq 2)$ variables on $n$ (sufficiently large) time points. For example, in case
of daily financial data we would expect to observe the price of all $p$ stocks on a particular day. This kind of data is called \emph{synchronously} observed data. On the other hand, if we don't have observations for one or several variables (or stocks) on a particular time point then we call it \emph{nonsynchronous /asynchronous} data. An example of such data is intraday stock price data. Within a particular day we can not expect to observe transactions in all stocks simultaneously. In Figure \ref{Figure 1}, we have shown the transaction/arrival times of two stocks within a small time interval.

The effect of asynchronicity can be quite serious on the estimation of model parameters. One of such phenomenon, reported by \citet{epps1979comovements}, is called Epps effect. Empirical results reported in that paper showed that the realized covariance between stock returns decreases as sampling frequency increases. Later the same phenomenon has been reported in several other studies on different stock markets, see \citet{zebedee2009closer} and foreign exchange market, see \citet{muthuswamy2001time}. It is also empirically shown , see \citet{reno2003closer}, that taking into account only the synchronous, or nearly synchronous, alleviates this underestimation problem.

Several studies have been devoted to the estimation of the covariance from intraday data. \citet{mancino2011estimating} analysed the performance of the \textit{Fourier} estimator originally proposed by \citet{malliavin2009fourier}. \citet{peluso2014bayesian} adopted a Bayesian Dynamic Linear model and treated asynchronous trading as missing observations in an otherwise synchronous series. \cite{corsi2012realized} proposed two covariance estimators, adapted to the case of rounding in the price time stamps, which can be seen as a general way of computing the Hayashi-Yoshida estimator (see \citet{hayashi2005covariance}) at a lower frequency. \cite{ZhangL(2011)} proposed a method called two-scale realized covariance estimator (TSCV) which combines two-scale sub-sampling and previous tick method that can simultaneously remove the bias due to microstructure noise and asynchronicity. \cite{fan2012vast} studies TSCV under high-dimensional setting. \cite{ait2010high} proposed quasi–maximum likelihood estimator of the quadratic covariance.

The correlation coefficient only captures the linear dependence. In this work, we aspire to focus on estimation of non-linear dependence structure through copula. Apart from modelling the complete dependence structure, one of many other advantages of copula is the flexibility it offers to model the complex relationship between variables in a rather simple manner. It allows us to model the marginal distributions as necessary and takes care of the dependence structure separately. It is also one of the most important tools to model tail dependence, which is the probability of extremely large or small return on one asset given that the other asset yielded an extremely large or small return, see \citet{xu2008estimating}. For this reason, copula is also a useful tool for modelling the joint distribution of default times. Therefore it is important for pricing Credit Spread Basket, Credit Debt Obligation, First to Default, $N$-to Default and other Credit derivative baskets, see \citet{malgrat2013pricing}. They also shows how copula helps to correlate the systematic risk to idiosyncratic risk. \citet{zhang2016copula} developed a class of copula structured multivariate maxima and moving maxima processes which is a flexible and statistically workable model to characterize joint extremes.   \\

In this paper, we will elaborately discuss the impact of asynchronicity on several measures of association in a general class of copulas.  We explain why there is a serious underestimation of the measures of association unless treated carefully. We propose an alternative method for the estimation of correlation. Moreover we prescribed methods for accurate estimation of the associated copula. It is also shown that the estimation of some commonly used measures of associations, like Kendall's tau, is challenging unless the underlying copula is determined. The rest of the paper  is  organized  as  follows. In  section  \ref{sec:Elliptical}  we  deal  with  the  elliptical  copula  parameter estimation  for  nonsynchronous  data  and  prove  the  main  theorems. Section \ref{sec:General} deals with a more general class of copulas.  In section \ref{sec:Simulation} and \ref{sec:Data} the results of simulation and real data analysis are shown.  We present the conclusions in section \ref{sec:Conclusions}. All the proofs are given in appendix \ref{Proofsec}.

\section{Estimation of Elliptical Copula}\label{sec:Elliptical}

Suppose there are two stocks and their log-prices at time $t\in(0,T)$ are denoted by $X_{t}$ and $Y_{t}$. By $R_{t}^{1}$ and $R_{t}^{2}$ we denote the corresponding log-returns. Although in the ideal world of the Black-Scholes model, the log returns are assumed to follow a Gaussian distribution, the stylized facts about financial market suggest that a distribution with a heavier tail needs to be considered. In the multivariate scenario, the search for such a model is challenging. In such situations copula appears to be a central tool at our disposal.

In Section \ref{sec:Simulation}, the results of a simulation study are reported where the effect of asynchronicity on the estimation of the correlation coefficient has been shown. The simulation results display severe underestimation. Before attempting to understand the problem and propose a remedy, we will present an algorithm to synchronize the data to make it suitable for standard multivariate analysis. We should note that some studies (see \citet{hayashi2005covariance}, \citet{buccheri2020score}) attempt to calculate integrated covariance without synchronizing the data.

\subsection{Pairing Method}\label{sec:method}
The prices of the stocks are observed at random times when transactions take place.
As a transaction in one stock wouldn't influence the transaction time in the other, it is reasonable to assume that the observation times of the two stocks are independent
Point processes. Therefore, if we have log prices of the first stock along with its time of occurrence as $(X_{i},t_{i}^{1}), i=1,2,..,n_{1}$ and that of the second stock as $(Y_{j},t_{j}^{2}), j=1,2,...,n_{2}$, then $t_{i}^{1}$s and $t_{j}^{2}$s are independent. Here $n_{1}$ and $n_{2}$ are the number of observations of first and second stock respectively, available on a particular day.

Before fitting a copula model, the observations of two stock prices need to be paired such that they can be treated as synchronously observed. The conventional synchronizing methods require a set (or sample) of $n$ time points $\tau_i ,i=1(1)n$ at which we would like to \textit{observe} a synchronized pair. For each stock, the tick information observed just previous to each such sampled time point $\tau_i$ is chosen to construct the synchronized pair ($X_{\tau_i},Y_{\tau_i}$), yielding $n$ such pairs.

It is evident from the above discussion that the number of synchronized pairs is less than both $n_{1}$ and $n_{2}$, unless we allow repetition. This means many observations in each stock will be removed and not to be used for further analysis. \textit{Generalized sampling times} are defined as the following.

\begin{defn} Suppose we have $M$ stocks. $t_{k}^{i}$ is the $k$-th arrival time of the $i$-th asset. Then $\{\tau_{j}$: $1\leq j\leq n\}$, are called generalized sampling times \cite{ait2005often} if
a) $0=\tau_{0}<\tau_{1}<...<\tau_{n}=T$.\\
b) $(\tau_{j-1},\tau_{j}]\cap\{t_{i,k}:k=1,...,n_{i}\}\neq\varnothing$ for some $i=1,...,M$.\\
c)$\mathrm{max}_{1\leq j\leq n}\delta_{j}\rightarrow0$ in probability, where $\delta_{j}=\tau_{j}-\tau_{j-1}$.
\end{defn}
If $\delta_{j}=\delta$, then it is called\textit{ Previous tick sampling}. In the above-mentioned method, an observation is uprooted from its original time point and assigned  to a sampled time point $\tau_{j}$, for some $j$. In contrast, we want to retain the actual times of the prices that are chosen to be paired. In other words, instead of having a pair like $(X_{\tau_{j}},Y_{\tau_{j}})$, we want to have a pair $(X_{t_{k_{i}^{1}}^{1}},Y_{t_{k_{i}^{2}}^{2}})$ where
$t_{k_{i}^{1}}^{1}\text{and }t_{k_{i}^{2}}^{2}$ are the times at which the $i$-th pair of stock-prices were observed. To emphasise this, we call the algorithm as the `\emph{pairing method}' (in contrast to `\emph{synchronizing method}').
The pairing method, to be followed throughout in this paper, is described through the following algorithm ($A_0$):
\begin{center}
\fbox{\begin{minipage}{0.8\textwidth}
\textbf{Algorithm} ($A_{0}$):

1. Take $i=1,$ $k_{i}^{1}=1$ and $k_{i}^{2}=1$

2. While $k^1_i \leq n_1$ and $k^2_i \leq n_2$:
\begin{itemize}
    \item If $t_{k_{i}^{2}}^{2}>t_{k_{i}^{1}}^{1}$ then find $m=\mathrm{max}\{j:\ t_{j}^{1}<t_{k_{i}^{2}}^{2}\}$.
The $i$th pair will be $(X_{t_{m}^{1}},Y_{t_{k_{i}^{2}}^{2}})$.
Modify $k_{i}^{1}=m$.
    \item If $t_{k_{i}^{2}}^{2}<t_{k_{i}^{1}}^{1}$ then find $m=\mathrm{max}\{j:\ t_{j}^{2}<t_{k_{i}^{1}}^{1}\}$.
The $i$th pair will be $(X_{t_{k_{i}^{1}}^{1}},Y_{t_{m}^{2}})$.
Modify $k_{i}^{2}=m$
    \item Modify $i=i+1$. $k_{i}^{1}=k_{i}^{1}+1$ and $k_{i}^{2}=k_{i}^{2}+1$.
\end{itemize}
\end{minipage} }
\end{center}

The pairs created by this algorithm are identical to the pairs created by "refresh time sampling" (see  \citet{barndorff2011multivariate}) but accommodates more information by retaining the transaction times. Instead of writing $(X_{t_{k_{i}^{1}}^{1}},Y_{t_{k_{i}^{2}}^{2}})$
we shall henceforth write $(X_{t(k_{i}^{1})},Y_{t(k_{i}^{2})})$.

\begin{figure}
\centering\includegraphics[scale=0.5]{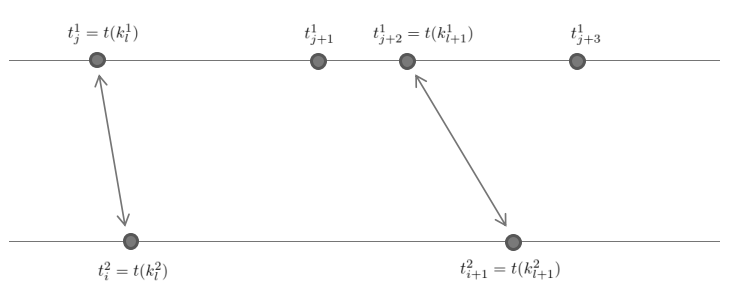}\caption{\label{algofig1}Two consecutive pairs represented by the arrows.}
\end{figure}
\begin{figure}
\centering\includegraphics[scale=0.5]{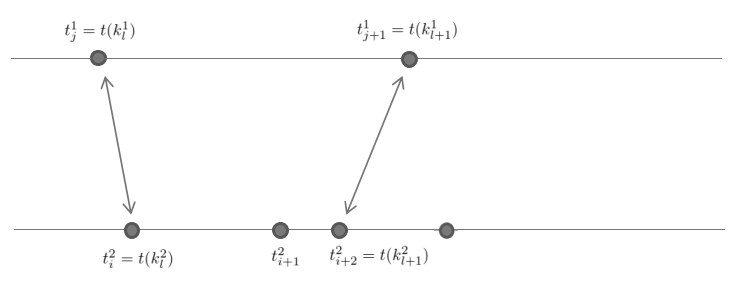}\caption{\label{algofig2} Two consecutive pairs represented by the arrows. }
\end{figure}

In Figure \ref{algofig1} and \ref{algofig2}, $t_{j}^{1}$
and $t_{i}^{2}$ are paired together as ($t(k_{l}^{1}),t(k_{l}^{2})$).
The figures illustrate how the next pair is going to be chosen using the algorithm. In figure \ref{algofig1}, $t_{j+1}^{1}<t_{i+1}^{2}$.
So $t_{i+1}^{2}=t(k_{l+1}^{2})$ and $t(k_{l+1}^{1})$ is chosen to be the largest of the arrival times in the first stock that are less than $t_{i+1}^{2}$. In figure \ref{algofig2}, $t_{j+1}^{1}>t_{i+1}^{2}$. So $t_{j+1}^{1}=t(k_{l+1}^{1})$ and $t(k_{l+1}^{2})$ is chosen to be the largest of the arrival times in the first stock that are less than $t_{j+1}^{1}$. The pairs are represented by the arrows.

\subsection{Estimation of Correlation coefficient}\label{sec:est}

Once we have paired observations, we can proceed
to calculate the correlation coefficient. We denote $(X,Y)$ as a bivariate rv with mean $0$ , variances $1$
and correlation $\rho$. $(X,Y)$ is independent of the arrival processes. Since the log returns $X_{t(k_{i}^{1})}-X_{t(k_{i-1}^{1})}$ and $Y_{t(k_{i}^{2})}-Y_{t(k_{i-1}^{2})}$ are calculated over two nonidentical time-intervals, namely $(t(k_{i}^{1}),t(k_{i}^{2}))$ and $(t(k_{i-1}^{1}),t(k_{i-1}^{2}))$, the correlation between the returns is heavily dependent on the length of overlapping and non-overlapping portions of these two time-intervals. To see this, first suppose $X_{t(k_{i}^{1})}-X_{t(k_{i-1}^{1})}=\sum_{i=m}^{l}(X_{t_{i+1}}-X_{t_{i}})$
for some $m$ and $l$. Here $\{t_{i}:i=1(1)(n_{1}+n_{2})\}$
is the set of combined (ordered) time points at which a transaction (in any
of the stocks) is noted. Then one of these four configurations is
true:

\begin{equation}\left[
\begin{array}{rcl}1. \quad  Y_{t(k_{i}^{2})}-Y_{t(k_{i-1}^{2})}&=&\sum_{i=m+1}^{l-1}(Y_{t_{i+1}}-Y_{t_{i}})\\
2. \quad Y_{t(k_{i}^{2})}-Y_{t(k_{i-1}^{2})}&=&\sum_{i=m-1}^{l-1}(Y_{t_{i+1}}-Y_{t_{i}})\\
3. \quad Y_{t(k_{i}^{2})}-Y_{t(k_{i-1}^{2})}&=&\sum_{i=m+1}^{l+1}(Y_{t_{i+1}}-Y_{t_{i}})\\
4. \quad  Y_{t(k_{i}^{2})}-Y_{t(k_{i-1}^{2})}&=&\sum_{i=m-1}^{l+1}(Y_{t_{i+1}}-Y_{t_{i}})\end{array}\right]\label{eqn:config}
\end{equation}

\begin{center}
\begin{figure}
\includegraphics[scale=0.5]{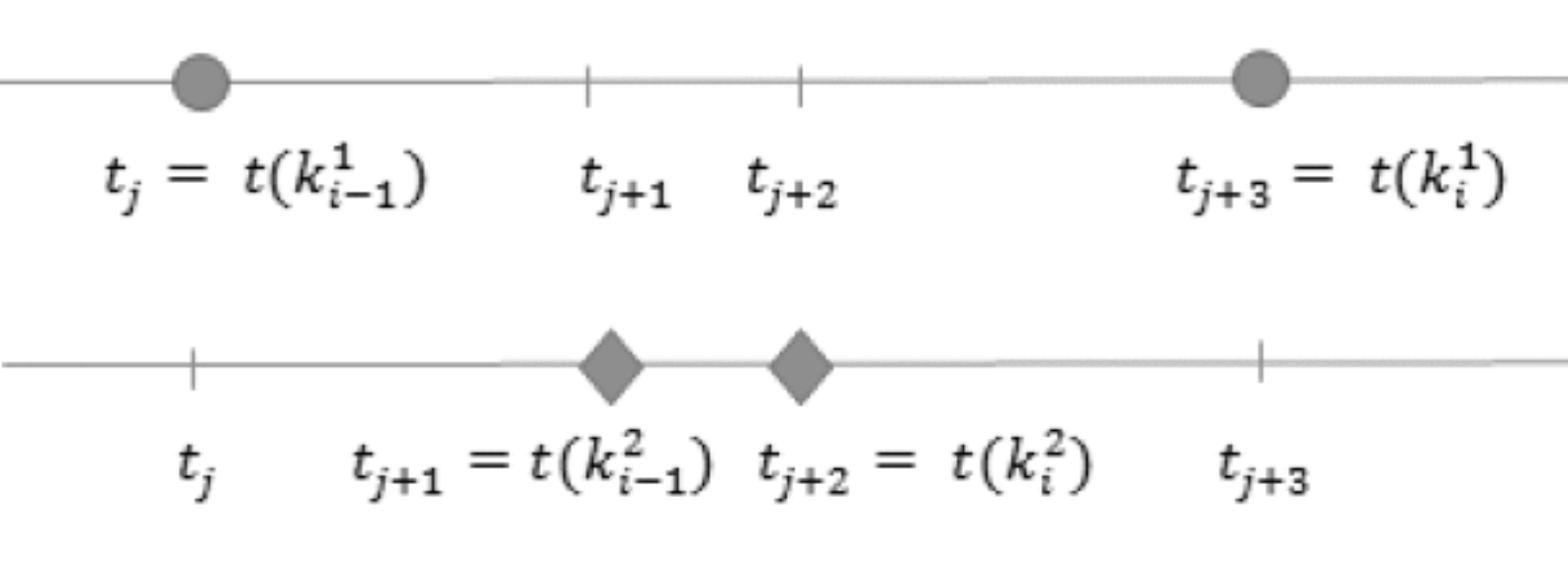}
\caption{Two consecutive pairs of log-returns are $(X_{t(k_{i-1}^{1})},Y_{t(k_{i-1}^{2})})$
and $(X_{t(k_{i}^{1})},Y_{t(k_{i}^{2})})$ with their corresponding
transaction times $(t(k_{i-1}^{1}),t(k_{i-1}^{2}))$, $(t(k_{i}^{1}),t(k_{i}^{2}))$}\label{Figure2}
\end{figure}
\end{center}
\begin{center}
\begin{figure}
\includegraphics[scale=0.5]{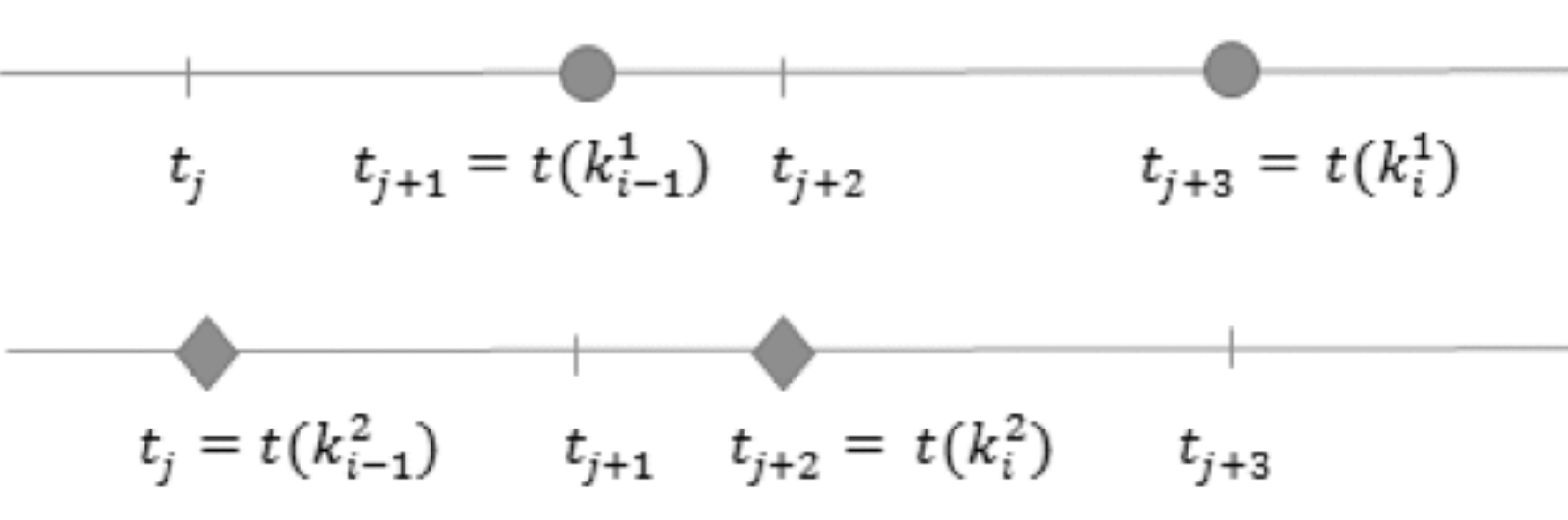}\caption{Two consecutive pairs of log-returns are $(X_{t(k_{i}^{1})},Y_{t(k_{i}^{2})})$
and $(X_{t(k_{i-1}^{1})},Y_{t(k_{i-1}^{2})})$ with their corresponding
transaction times $(t(k_{i}^{1}),t(k_{i}^{2}))$, $(t(k_{i+1}^{1}),t(k_{i+1}^{2}))$}\label{Figure3}
\end{figure}
\end{center}

See Figure \ref{Figure2} and \ref{Figure3} for illustrations of the first two configurations where two consecutive
pairs of log-prices are $(X_{t(k_{i}^{1})},Y_{t(k_{i-1}^{2})})$ and
$(X_{t(k_{i}^{1})},Y_{t(k_{i}^{2})})$ with their corresponding transaction
times $(t(k_{i-1}^{1}),t(k_{i-1}^{2}))$, $(t(k_{i}^{1}),t(k_{i}^{2}))$.

We define a random variable $I_i$, denoting the overlapping
time interval of $i$th interarrivals corresponding to $X_{t(k_{i}^{1})}-X_{t(k_{i-1}^{2})}$
and $Y_{t(k_{i}^{2})}-Y_{t(k_{i-1}^{2})}$
\[
I_i=\begin{cases}
t(k_{i}^{2})-t(k_{i-1}^{2}) & \mathrm{if}\ Y_{t(k_{i}^{2})}-Y_{t(k_{i-1}^{1})}=\sum_{i=m+1}^{l-1}(Y_{t_{i+1}}-Y_{t_{i}})\\
t(k_{i}^{2})-t(k_{i-1}^{1}) & \mathrm{if}\ Y_{t(k_{i}^{2})}-Y_{t(k_{i-1}^{1})}=\sum_{i=m-1}^{l-1}(Y_{t_{i+1}}-Y_{t_{i}})\\
t(k_{i}^{1})-t(k_{i-1}^{2}) & \mathrm{if}\ Y_{t(k_{i}^{2})}-Y_{t(k_{i-1}^{1})}=\sum_{i=m+1}^{l+1}(Y_{t_{i+1}}-Y_{t_{i}})\\
t(k_{i}^{1})-t(k_{i-1}^{1}) & \mathrm{if}\ Y_{t(k_{i}^{2})}-Y_{t(k_{i-1}^{1})}=\sum_{i=m-1}^{l+1}(Y_{t_{i+1}}-Y_{t_{i}})
\end{cases}
\]

For Figure \ref{Figure2}, $I_i=t(k_{i}^{2})-t(k_{i-1}^{2})$ and for Figure \ref{Figure3}, $I_i=t(k_{i}^{2})-t(k_{i-1}^{1})$.

First consider Figure \ref{Figure2}. Define $\mathbb{T}=\{t_{i}:i=1(1)n\}$ with $n=n_{1}+n_{2}$, where $n_{1}$, $n_{2}$ are the number of observations in the two stocks. The conditional expectation:

\begin{align*} & E\big[(X_{t(k_{i}^{1})}-X_{t(k_{i-1}^{1})})(Y_{t(k_{i}^{2})}-Y_{t(k_{i-1}^{2})})| \mathbb{T}\big]\\ & =E\big[(X_{t(k_{i}^{1})}-X_{t(k_{i-1}^{1})})(Y_{t(k_{i}^{2})}-Y_{t(k_{i-1}^{2})})|\mathbb{T},\ (t(k_{i-1}^{1}),t(k_{i-1}^{2}),t(k_{i}^{1}),t(k_{i}^{2}))=(t_{j},t_{j+1},t_{j+3},t_{j+2})\big]\\ & =E\big[(X_{t_{j+3}}-X_{t_{j+2}}+X_{t_{j+2}}-X_{t_{j+1}}+X_{t_{j+1}}-X_{t_{j}})(Y_{t_{j+2}}-Y_{t_{j+1}})|(t_{j},t_{j+1},t_{j+3},t_{j+2})\big]\\ & =E\big[(X_{t_{j+2}}-X_{t_{j+1}})(Y_{t_{j+2}}-Y_{t_{j+1}})|(t_{j},t_{j+1},t_{j+3},t_{j+2})\big]\\ & =E(X\sqrt{t_{j+2}-t_{j+1}}.Y\sqrt{t_{j+2}-t_{j+1}}|(t_{j},t_{j+1},t_{j+3},t_{j+2}))\\ & =(t_{j+2}-t_{j+1})E(XY)\\ & =(t_{j+2}-t_{j+1})\rho \end{align*} Thus, $E[(X_{t(k_{i}^{1})}-X_{t(k_{i-1}^{1})})(Y_{t(k_{i}^{2})}-Y_{t(k_{i-1}^{2})})]=\rho E(t(k_{i}^{2})-t(k_{i-1}^{2}))$ \begin{align*} \textrm{Also,}\quad E[(X_{t(k_{i}^{1})}-X_{t(k_{i-1}^{1})})^{2}|\mathbb{T}] & =E(X_{t_{i+3}}-X_{t_{i}}|t_{j},t_{j+3})^{2}\\ & E(X_{t_{j+3}}-X_{t_{j}}|t_{j},t_{j+3})^{2}\\ & =(t_{j+3}-t_{j})EX{}^{2}\\ & =t_{j+3}-t_{j} \end{align*}
and hence, $E(X_{t(k_{i}^{1})}-X_{t(k_{i-1}^{1})})^{2}=E(t(k_{i}^{1})-t(k_{i-1}^{1}))$.
Similarly, $E(Y_{t(k_{i}^{2})}-Y_{t(k_{i-1}^{2})})^{2}=E(t(k_{i}^{2})-t(k_{i-1}^{2}))$.
Thus for the case of Figure \ref{Figure2}, \[Cor[(X_{t(k_{i}^{1})}-X_{t(k_{i-1}^{1})}),(Y_{t(k_{i}^{2})}-Y_{t(k_{i-1}^{2})})]=\frac{\rho E(t(k_{i}^{2})-t(k_{i-1}^{2}))}{\sqrt{E(t(k_{i}^{1})-t(k_{i-1}^{1}))}.\sqrt{E(t(k_{i}^{2})-t(k_{i-1}^{2}))}} \]

These examples lead us to our first theorem.

We consider the following assumptions ($\mathcal{A}$):

$\mathcal{A}_{1}$: The log return process follows independent and
stationary increment property

$\mathcal{A}_{2}$: The observation times (arrival process) of two
stocks are independent Renewal processes and $n\rightarrow\infty$
as $n_{1},n_{2}\rightarrow$ $\infty$

$\mathcal{A}_{3}$: Estimation is based on paired data obtained by
algorithm $A_0$

\begin{thm}
\textup{\label{thcoppara}} Under the assumptions $\mathcal{A}_{1}-\mathcal{A}_{3},$ $\hat{\theta}$
defined below is a consistent estimator of the true correlation coefficient
\[ \hat{\theta}=\hat{\rho}\frac{\sqrt{m_{1}.m_{2}}}{m(I)} \] where
for $l=1,2$ $m_{l}=\frac{1}{n}\sum_{i=1}^{n}(t(k_{i}^{l})-t(k_{i-1}^{l})),$$m(I)=\frac{1}{n}\sum_{i=1}^{n}I_{i}$
and $\hat{\rho}$ being the sample mean and sample correlation coefficient
based on the pairs.

Moreover,
\[
\sqrt{n}(f(\hat{\theta})-f(\theta))\stackrel{d}{\rightarrow}N(0,1)
\]
where $f(\theta)=\frac{1}{2}[log(1+\frac{\theta}{\gamma})-log(1-\frac{\theta}{\gamma})]$
and $f(\hat{\theta})=\frac{1}{2}[log(1+\frac{\hat{\theta}}{\gamma})-log(1-\frac{\hat{\theta}}{\gamma})]$
and
\[
\gamma=\frac{\sqrt{E(t(k_{i}^{1})-t(k_{i-1}^{1}))E(t(k_{i}^{2})-t(k_{i-1}^{2}))}}{E(I_{i})}
\]
\end{thm}

Proof of Theorem \ref{thcoppara} is given in Appendix \ref{Proofsec}.

According to this theorem in order to get a consistent estimator, we
need to multiply the usual sample correlation coefficient, based on
the paired observations (by algorithm $A_{0}$), by a correction factor.
The correction factor is a function of only $t_{k_{i}^{1}}$ and
$t_{k_{i}^{2}}$ for $i=1(1)n$ i.e. it is only dependent on the arrival process and not on the copula.

\subsection{Nonlinear dependence and Elliptical copula} So far we were dealing with linear dependence through the correlation coefficient. In this section we will deal with nonlinear dependence through copula. We will restrict our attention to elliptical copula. The \emph{Gaussian copula} is the most widely used elliptical copula which mimics the dependence structure of a multivariate Gaussian distribution. But it does not capture the nonlinear dependence. It is well-known that a Gaussian copula with correlation coefficient zero reduces to independent copula. But this is not true in general. For example in case of another common elliptical copula, namely the \textit{t} copula, the parameter captures the linear dependence but the form of the copula function accommodates for nonlinear dependence. We will now discuss the effect of asynchronicity on copula estimation.

By Sklar's theorem (see \citet{nelsen2007introduction}), the distribution function of the log returns $R^1$ and $R^2$ can be expressed as $F(r^{1},r^{2})=C(F_{1}(r^{1}),F_{2}(r^{2});\theta)$, where $C$ is the unique copula associated with $F$. Asynchronicity not only affects the estimation of $\theta$, but also the estimation of the copula function because $r^{1}$ and $r^{2}$ are assumed to be observed synchronously. The convergence of $\hat{C}(\hat{F}_{1}(r^{1}),\hat{F}_{2}(r^{2});\hat{\theta})$,
where $\hat{F}_{1}(.)$ and $\hat{F}_{2}(.)$ are empirical distribution functions of $F_{1}(.)$ and $F_{2}(.)$, needs more than the convergence of $\hat{\theta}$. The next theorem tries to address this concern. Before stating the theorem we will make an additional assumption which ensures that the probability of both the missing value of return at $t_{k_{i}^{1}}$ and observed value of return at $t_{k_{i}^{2}}$ lying in an interval of length $2\delta$ is in the order of $\frac{\delta}{n^{\psi}}$ for $\psi>0$. Define, $R^1_t(k^l_i)=X_t(k^l_i)-X_t(k^l_i)$ and $R^2_t(k^l_i)=Y_t(k^l_i)-Y_t(k^l_i)$ for $l=1,2$.\\
$\mathcal{A}_{4}:$ $P[|R_{t(k^1_i)}^l-R_{t(k^2_i)}^l|\leq 2\delta]=O(\frac{\delta}{n^{\psi}})$
for $l=1,2$ where $|\delta|=|\mathrm{max_{i}}(r_{t(k_{i}^{1})}-r_{t(k_{i}^{2})})|<M$
with $\psi>0$ and $M$ being a positive real number.

\begin{thm}
\textup{\label{thcopfn}} If the true underlying copula is an elliptical copula then under $\mathcal{A}_{1}-\mathcal{A}_{4}$, $C(\hat{F}_{1}(r_{1}),\tilde{F}_{2}(r_{2});\hat{\theta})$ is uniformly convergent to the true copula, where
$\hat{F}_{1}(.)$ and $\tilde{F}_{2}(.)$ are the empirical distribution functions of the marginals of $R^{1}$ and $R^{2}$ computed from the paired data and $\hat{\theta}$ is defined as in Theorem \ref{thcoppara}.
\end{thm}

Proof of Theorem \ref{thcopfn} is given in Appendix \ref{Proofsec}.

\subsection{Expected loss of data and $\gamma$:} Recall that the correction factor in Theorem \ref{thcoppara} is a function of the arrival process only. It is worthwhile to express $\gamma$ in terms of the underlying parameters of the arrival processes. In the next theorem, we will try to do so. But the implication of the theorem
goes beyond this purpose. Remember that all the synchronization methods
we discussed have one problem. It results in loss of data, which is evident from Figure \ref{algofig1}. The second  observation
of the first stock will not be included in any of the pairs and therefore will be wasted. So one can ask the question that what proportion of observations (of each stock)
will be wasted by using our pairing method ($A_{0}$). This can be answered if we can compare average interarrival
length in a stock (for example $E(t_{i}^{1}-t_{i-1}^{1})$ for the
first stock) and average interarrival length formed by the pairs ($E(t_{k_{i}}^{1}-t_{k_{i-1}}^{1})$
for the first stock). One important point to note here is that even if the two initial point processes ${t_{i}^{1}}:i=1(1)n_{1}$ and ${t_{i}^{2}}:i=1(1)n_{2}$ are independent, the point processes after pairing the observations- ${t_{k_{i}}^{1}}:i=1(1)n$ and ${t_{k_{i}}^{2}}:i=1(1)n$- are not independent. This is due to the fact that the pairing method ($A_{0}$) involves arrivals of both the stocks. Due to that fact, we will see in the next theorem both $E(t_{k_{i}}^{1}-t_{k_{i-1}}^{1})$ and $E(t_{k_{i}}^{2}-t_{k_{i-1}}^{2})$ involves $\lambda_{1}, \lambda_{2}$, the parameters of the two point processes.

\begin{thm}
\textup{\label{thdataloss}}Suppose the two underlying point processes
are Poisson processes with parameters $\lambda_{1}$ and $\lambda_{2}$
and $Z_{k}\sim Beta(1,k)$ for $k>1$ then,
\begin{enumerate}[label=(\alph*)]
    \item
        $E(I) = \frac{1}{2}\sum_{n=1}^{\infty}n[(\frac{1}{\lambda_{1}}+\frac{1}{\lambda_{2}})(p_{n}+q_{n})]$

    \item
        $E(t_{k_{i}}^{1}-t_{k_{i-1}}^{1}) = \frac{\eta_{1}}{\lambda_{1}}$

    \item $E(t_{k_{i}}^{2}-t_{k_{i-1}}^{2}) = \frac{\eta_{2}}{\lambda_{2}}$
\end{enumerate}

where
\[
p_{n}=F_{Z_{n+1}}(\frac{\lambda_{2}}{\lambda_{1}+\lambda_{2}})-F_{Z_{n}}(\frac{\lambda_{2}}{\lambda_{1}+\lambda_{2}}),
\]

\[
q_{n}=F_{Z_{n+1}}(\frac{\lambda_{1}}{\lambda_{1}+\lambda_{2}})-F_{Z_{n}}(\frac{\lambda_{1}}{\lambda_{1}+\lambda_{2}})
\]

and for i=1,2
\[
\eta_{i}=\sum_{k=1}^{\infty}[\{F_{B(1,k+1)}(1-\frac{\lambda_{i}}{\lambda_{1}+\lambda_{2}})-F_{B(1,k)}(1-\frac{\lambda_{i}}{\lambda_{1}+\lambda_{2}})\}+k\{F_{B(1,k+1)}(\frac{\lambda_{i}}{\lambda_{1}+\lambda_{2}})-F_{B(1,k)}(\frac{\lambda_{i}}{\lambda_{1}+\lambda_{2}})\}]
\]
\end{thm}

Proof of Theorem \ref{thdataloss} is given in Appendix \ref{Proofsec}.

As a consequence of this theorem we have-
\begin{equation}
\gamma=\frac{\sqrt{\frac{\eta_{1}\eta_{2}}{\lambda_{1}\lambda_{2}}}}{\frac{1}{2}\sum_{n=1}^{\infty}n[(\frac{1}{\lambda_{1}}+\frac{1}{\lambda_{2}})(p_{n}+q_{n})]}
\end{equation}

\section{Extension to general copula }\label{sec:General}

In this section, we will deal with a more general
class of copulas. As the argument in Section \ref{sec:Elliptical} is entirely based
on the correlation coefficient it can not be directly extended to a larger class of copulas. This is precisely because for a general copula there is no direct relation between the Pearson's correlation coefficient and the copula parameter.

We propose to use Kendall's Tau to capture the copula dependence. The definition of Kendall's tau is
\begin{equation*}
    \rho_{\tau}(X,Y) := E(\mathrm{sign}((X-\tilde{X})(Y-\tilde{Y}))),
\end{equation*}
where $\tilde{X}$ and $\tilde{Y}$ are identical but independent
copies of $X$ and $Y$. The relation between Kendall's Tau and the copula is captured through the following equation.
\begin{equation}
    \rho_{\tau}(X,Y)=4\int_{0}^{1}\int_{0}^{1}C(u_{1},u_{2})dC(u_{1},u_{2})-1.
\end{equation}
If $X$ and $Y$ be random variables with an Archimedean copula $C$ generated by $\phi$ in $\Omega$ then
\begin{equation}
\rho_{\tau}(X,Y)=1+4\int_{0}^{1}\frac{\phi(t)}{\phi'(t)}dt.
\end{equation}
For the elliptical copulas a simplified form can be derived,
\begin{equation}
\rho_{\tau}(X,Y)=\frac{2}{\pi}\mathrm{arcsin}\rho\label{eq:elliptical_para_tau_relation}.
\end{equation}
So we can study how Kendall's tau is affected by asynchronicity. Thereupon we will gauge the impact on the copula parameter using the above mentioned relation.

\begin{figure}
\includegraphics[scale=0.5]{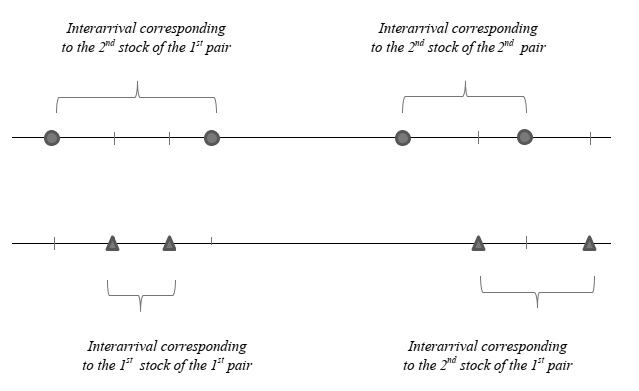}\caption{Two independent observations (pairs) of two different configurations.}\label{piars_diff_config1}
\end{figure}

\subsection{Underestimation of Kendall's tau}
\begin{figure}
\includegraphics[scale=0.6]{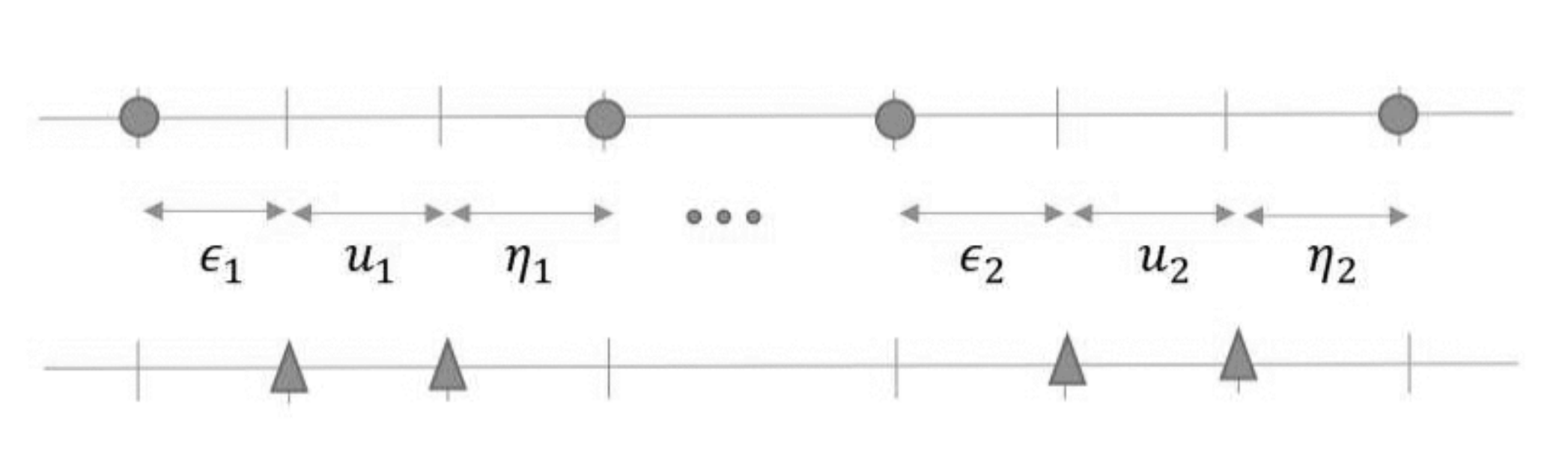}\caption{An example of arrival times of two pairs with same configuration}\label{Figure4}
\end{figure}

 The problem with nonsynchronous data is that any two independent pairs of returns can not be taken as identical copies of each other. To see this, consider figure \ref{piars_diff_config1}, where arrival times of the 1st stock are denoted by triangles and arrival times of the second stock are denoted by circles. After applying the pairing method, suppose the first circle and first triangle represent the location of the first pair of prices. Similarly, the second circle and the second triangle represent the next pair. From the figure, it is evident that these two pairs are forming an example of the second configuration (see eq. \ref{eqn:config}). Similarly, the 3rd and 4th pair constitutes an example of the 4th configuration. So the corresponding returns may not be considered as identically distributed. In this subsection, we will measure the Kendall's tau using only the returns with same configuration. Figure \ref{Figure4} represents the arrival times of two pairs of the same configuration.

As illustrated in Figure \ref{Figure4}, suppose we have two non-overlapping inter-arrivals $u_{1}$ and $u_{2}$ for the first stock and $\epsilon_{1}+u_{1}+\eta_{1}$ and $\epsilon_{2}+u_{2}+\eta_{2}$ for the second stock, with arrival times denoted by the triangles and circles respectively. The log returns corresponding to the inter-arrivals
of the first stock are given by $R^{1}_1=R^{1}(u_{1})$ and $R^1_2=R^{1}(u_{2})$.
Similarly, the log returns corresponding to the intervals of the second stock are denoted by $R^2_{1}=R^2(\epsilon_{1}+u_{1}+\eta_{1})=R^2(\epsilon_{1})+R^2(u_{1})+R^2(\eta_{1})$
(due to independent increment property) and $R^2_{2}=R^2(\epsilon_{2}+u_{2}+\eta_{2})=R^2(\epsilon_{2})+R^2(u_{2})+R^2(\eta_{2})$.
In the following section, we will focus on the two specific configurations (eq. \eqref{eqn:config}).

Define, \[ A=(R^1(I_{1})-R^1(I_{2}))(R^2(I_{1})-R^2(I_{2})) \] and \[ B=\begin{cases} (R^1(I_{1})-R^1(I_{2}))(R^2(I_{1}^{c})-R^2(I_{2}^{c})) & \mathrm{for\ 4th\ configuration}\\ (R^1(I_{1}^{c})-R^1(I_{2}^{c}))(R^2(I_{1})-R^2(I_{2})) & \mathrm{for\ 1st\ configuration} \end{cases} \]

where $I_{i}$ and $I_{i}^{c}$ are respectively overlapping and non-overlapping regions of the $i$th pair of returns. In the above example, $\mathrm{length}(I_{1})=u_{1}$, $\mathrm{length}(I_{2})=u_{2}$, $\mathrm{length}(I_{1}^{c})=\epsilon_{1}+\eta_{1}$ and $\mathrm{length}(I_{2}^{c})=\epsilon_{2}+\eta_{2}$. Note that for the first and fourth configurations, $E(\mathrm{sign}(A))$ gives us true Kendall's tau. We cannot calculate $E(\mathrm{sign}(A))$ because $R^2(I_{1})$ and $R^2(I_{2})$ are not observed. Instead, we observe $R^2(I_{1}\cup I_{1}^c)$ and $R^2(I_{2}\cup I_{2}^{c})$. Therefore the observed Kendall's Tau is $E(\mathrm{sign}(A+B))$. In this section, we will try to find out the relation between $E(\mathrm{sign}(A))$ and $E(\mathrm{sign}(A+B))$.

In order to establish our result, we need some assumptions.
Suppose $X$ and $Y$ are positively associated random variables.
Let $(X_{1},Y_{1})$ and $(X_{2},Y_{2})$ be two identical copies
of $(X,Y)$. Then, given the information that $Y_{1}-Y_{2}>0$, we
would expect that $X_{1}-X_{2}$ is more likely to be positive than negative. Intuitively, positive association would also suggest that given the information $Y_{1}-Y_{2}\in S\subset\mathbb{R^{+}}$, $X_{1}-X_{2}$ is more likely to be positive. This notion is not in general captured by any known measure of association.  For each of the following we define $(X_{1},Y_{1})$ and $(X_{2},Y_{2})$ as two identical copies of $(X,Y)$, $U=X_{1}-X_{2}$ and $V=Y_{1}-Y_{2}$.

Assumptions($\mathcal{B}$) stated below, try to capture the above idea: \\
$\mathcal{B}_1$: If $P(UV>0)-P(UV<0)>0\ \mathrm{(or\ <0)}$
then for all $M>0$,\\ $P(U>0|0<V<M)-P(U<0|0<V<M)  \geq0\ \mathrm{(or\ <0)}$ and\\
$P(V>0|0<U<M)-P(V<0|0<U<M)  \geq0\ \mathrm{(or\ <0)}$.\\
$\mathcal{B}_2$: If  $P(UV>0)-P(UV<0)>0\ (or\ <0)$ then for all $M>0$,\\
$P(UV>0|\ |V|>M)-P(UV<0|\ |V|>M)  >0\ (or\ >0)$ and\\
$P(UV>0|\ |V|>M)-P(UV<0|\ |U|>M)  >0\ (or\ >0).$

Before stating the main theorems, we will first state some Lemmas which will help us to prove the theorems.
\begin{lem}
\textup{\textcolor{black}{\label{lemma1} }}$E(\mathrm{sign}(A)|\mathrm{sign}(A)\neq\mathrm{sign}(B))=E(\mathrm{sign}(A))$.
\end{lem}
Proof of Lemma \ref{lemma1} is given in Appendix \ref{Proofsec}.
\begin{lem}
\textup{\textcolor{black}{\label{lemma2}}}$E(\mathrm{sign}(A)|\mathrm{sign}(A)\neq\mathrm{sign}(B),|A|<|B|)=E(\mathrm{sign}(A)||A|<|B|)$
\end{lem}
This is a straightforward consequence of Lemma \ref{lemma1} and the independence of $\{\mathrm{sign}(A)\neq\mathrm{sign}(B)\}$ and $\{|A|<|B|\}$.
\begin{thm}
\label{1stth}Under the Assumption $\mathcal{B}$, for the pairs with 1st and 4th configuration, \[ |\tilde{\rho_{\tau}}|>|\rho_{\tau}| \] where $\rho_{\tau}$
is the Kendall's tau calculated on the paired data with 1st and 4th configurations, i.e. \textup{$\rho_{\tau}=E(\mathrm{sign}(X_{1}-X_{2})(Y_{1}-Y_{2}))$,}
where $(X_{1},Y_{1})$ and $(X_{2},Y_{2})$ are independent pairs
of the same configurations.\textup{ }
\end{thm}
Proof of Theorem \ref{1stth} is given in Appendix \ref{Proofsec}.
Now we will show that $\mathrm{sign}(\rho_{\tau})=\mathrm{sign}(\hat{\rho}_{\tau})$.
\begin{thm}
\label{oldth}For the pairs with 1st and 4th configuration, \[ \rho_{\tau}=E\big[\mathrm{sign}(A|\mathrm{sign}(A)\neq \mathrm{sign}(B)\&|A|>|B|)\big]P(|A|>|B|) \]

where $\rho_{\tau}$ is the Kendall's tau calculated on the paired
data with 1st and 4th configurations, i.e. \textup{$\rho_{\tau}=E(\mathrm{sign}(X_{1}-X_{2})(Y_{1}-Y_{2}))$,}
where $(X_{1},Y_{1})$ and $(X_{2},Y_{2})$ are independent pairs
of the same configurations.\textup{ }
\end{thm}
Proof of Theorem \ref{oldth} is given in Appendix \ref{Proofsec}.
By Lemma \ref{lemma2}, \[ \rho_{\tau}=E\big[\mathrm{sign}(A||A|>|B|)\big]P(|A|>|B|). \]
This together with Assumption $\mathcal{B}$ implies that, $\mathrm{sign}(\rho_{\tau})=\mathrm{sign}(\hat{\rho}_{\tau})$.

Theorems \ref{1stth} and \ref{oldth} together imply that the estimator of Kendall's tau obtained after pairing the observed asynchronous data underestimates the true parameter under the assumption $\mathcal{B}$ for the 1st and 4th configurations. Similar results can be established under the other two configurations.

\subsection{Corrected Estimator} 
\begin{figure}{}
    \centering
    \includegraphics[width=0.5\textwidth]{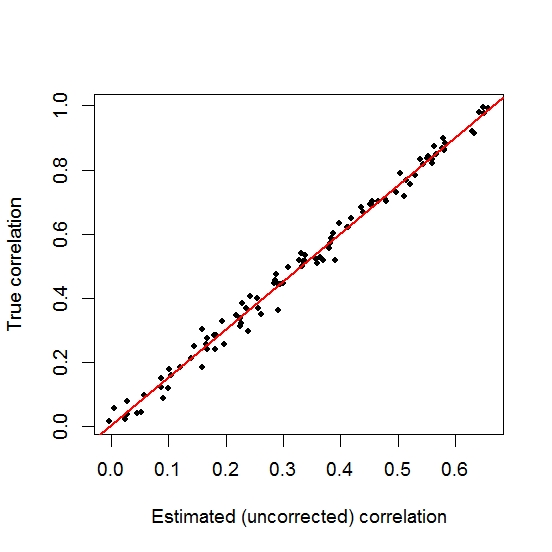}%
    \includegraphics[width=0.5\textwidth]{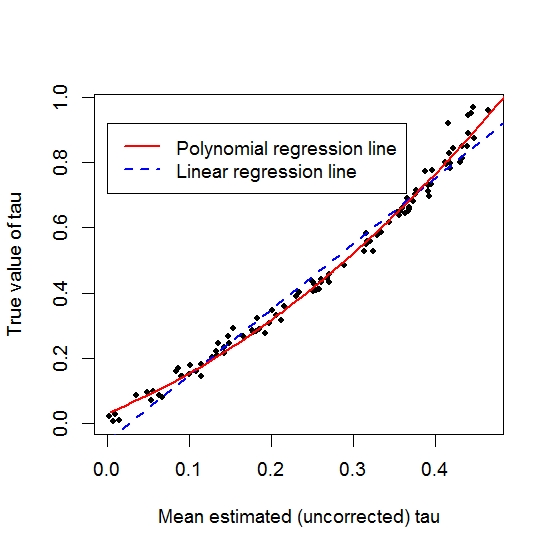}
    \caption{True vs mean estimated correlation obtained from 100 simulations for Gaussian copula (left) and true vs mean estimated Kendall's tau obtained from 100 simulations for Clayton copula (right). The margins in both the cases are taken to be Gaussian with 0 mean and 1 standard deviation.}
    \label{true vs estimated}
\end{figure}

Similar to section \ref{sec:Elliptical}, we would like to find a correction factor, that only depends on the arrival times, for a more general class of copula. For Elliptical copula, the value of the correction factor is not dependent on the value of the parameter. This is evident from Fig. \ref{true vs estimated} (left panel), showing the true and uncorrected mean estimated parameter of Gaussian copula for simulated nonsynchronous data. We generated the arrival times according to a pre-specified Poisson process. We can see that the true and estimated parameters lie along the regression line where the intercept term of the regression line is insignificant. This suggests that the corrected estimator should be a constant times the uncorrected one. This constant was the correction factor derived in Theorem \ref{thcoppara}.

On the other hand in the figure for Clayton copula (right panel, Fig. \ref{true vs estimated}),  we can see that a straight line would not be a good candidate to model the relation between the true and uncorrected estimated Kendall's tau. This means we should not aspire to find a simple multiplicative correction factor that would give us the value of the true parameter. On inspection, a second degree polynomial seems to be a good model. Bu the same procedure, a second-degree polynomial seems to be appropriate for the Gumbel copula as well. We therefore, use a quadratic model to obtain the corrected estimator. The detailed steps are outlined below:\\
\hrule
\begin{enumerate}
    \item From the observed data, estimate the two arrival processes independently.
    \item Estimate the univariate marginal distributions.
    \item Using the pairing algorithm described in section \ref{sec:est}, pair the observations.
    \item With paired data, we can now see which copula fits best to the data. It can be obtained through AIC or BIC criterion.
    \item Estimate the Kendall's tau (uncorrected) from this paired data.
    \item Prefix $K$ copula parameters (or equivalently Kendall's tau). For each parameter, with the information of the underlying copula, arrival processes, and marginals, we now simulate $N$ nonsynchronous samples (the technique of generating nonsynchronous data is discussed in section \ref{sec:Simulation}).
    \item For each sample, calculate uncorrected estimate and plot the estimates and the true Kendall's tau in a plot like Fig. \ref{Clayton CI} (right panel).
    \item Fit a suitable quadratic regression for such a plot.
    \item From the regression equation, find the corrected Kendall's tau corresponding to the estimated value of the Kendall's tau (obtained from step 5).
\end{enumerate}
\hrule

Note that the above procedure yields an interval estimator for Kendall's tau by considering the prediction interval in the regression. In section \ref{sec:Simulation} we study the coverage probability of such intervals through simulations and compare them to other interval estimates.

\section{Simulation}\label{sec:Simulation}
We simulated data of synchronized log-returns of two stocks for $n_1+n_2$ time points. The time points are generated by a Poisson Process. Corresponding $n_1+n_2$ returns are drawn randomly from a bivariate distribution determined by a pre-specified copula and margins. These $n_1+n_2$ pairs are then transformed appropriately to represent log-prices on the corresponding interarrivals. Now from the first stock, we randomly delete $n_2$ time points and their corresponding prices. The remaining $n_1$ data points constitute the data for the first stock. For the second stock, we keep the time points which were deleted from the first stock and delete rest of the time points. These time
points, along with their corresponding log-prices, constitute data for the second stock. So now we have nonsynchronous data for the two stocks.

\subsection{Estimation of copula parameter of Elliptical copulas}
\begin{figure}
\begin{minipage}[t]{0.45\columnwidth}%
\includegraphics[scale=0.4]{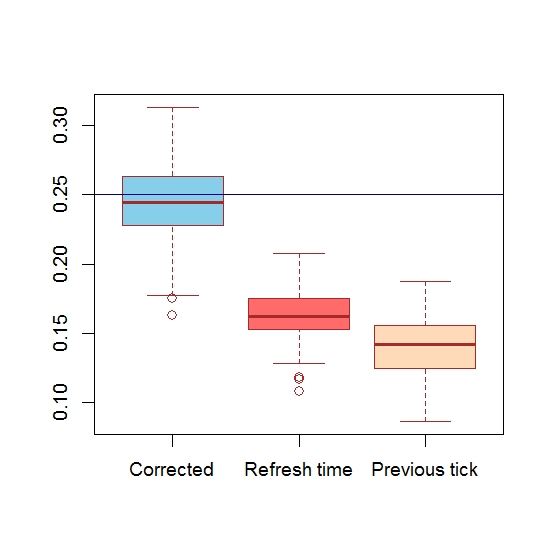}%
\end{minipage}\hfill{}%
\begin{minipage}[t]{0.45\columnwidth}%
\includegraphics[scale=0.4]{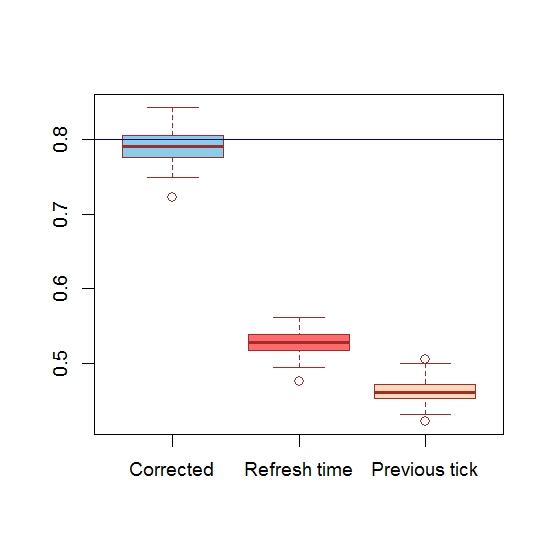}%
\end{minipage}\caption{Boxplots for $\rho=0.25$ (left) and $\rho=0.8$ (right);
the first boxplot is for the corrected estimates of 100 simulations with sample size 2000. The middle and the last plots correspond to the estimates from refresh time sampling and previous tick sampling respectively.}\label{Figure:boxplot}
\end{figure}

In the following simulation study, we test the performance of the method, prescribed in Theorem \ref{thcoppara}, to estimate the copula parameter. To do so, we first choose a Gaussian copula and generate 100 instances of nonsynchronous data by the method mentioned above. Initially, both $n_1$ and $n_2$ are taken to be the same. The mean, variance and Mean Square Error of the 100 estimates are reported in Table \ref{Table 1} and Table \ref{Table 2}. In Figure \ref{Figure:boxplot},
we show the boxplots for $\rho=0.8$. The boxplot on the left corresponds to the corrected estimate and those on the middle and right corresponds to uncorrected estimates obtained from refresh time sampling and previous tick sampling respectively. The horizontal line suggests the true parameter.

\begin{table} \begin{centering} \begin{tabular}{cccccc} \hline $\rho$ & $n$ & Estimate from & Estimates from & Corrected \tabularnewline & & previous tick sampling & refresh time sampling & estimate \tabularnewline \hline -0.4 & 800 & -0.2316 & -0.2680 & -0.4022 \tabularnewline & & (0.049)& (0.046)& (0.069) \tabularnewline \hline 0.1 & 800 & 0.061 & 0.069 & 0.1049 \tabularnewline & & (0.043)& (0.042)& (0.063) \tabularnewline \hline 0.2 & 800 & 0.1205 & 0.1237 & 0.1853 \tabularnewline & & (0.049)& (0.045)& (0.067)\tabularnewline \hline 0.8 & 800 & 0.4675 & 0.5255 & 0.7885 \tabularnewline & & (0.038)& (0.035)& (0.051)\tabularnewline \hline -0.4 & 2000 & -0.2325 & -0.2682 & -0.4022 \tabularnewline & & (0.034)& (0.025)& (0.039) \tabularnewline \hline 0.1 & 2000 & 0.056 & 0.065 & 0.098 \tabularnewline & & (0.033)& (0.026)& (0.039) \tabularnewline \hline 0.2 & 2000 & 0.1115 & 0.1274 & 0.1911 \tabularnewline & & (0.027)& (0.031)& (0.046)\tabularnewline \hline 0.8 & 2000 & 0.4613 & 0.5258 & 0.7888 \tabularnewline & & (0.028)& (0.019)& (0.029)\tabularnewline \hline -0.4 & 5000 & -0.2289 & -0.2637 & -0.3956 \tabularnewline & & (0.018)& (0.017)& (0.026) \tabularnewline \hline 0.1 & 5000 & -0.061 & 0.065 & 0.099 \tabularnewline & & (0.049)& (0.019)& (0.029) \tabularnewline \hline 0.2 & 5000 & 0.1165 & 0.1357 & 0.2036 \tabularnewline & & (0.019)& (0.015)& (0.023)\tabularnewline \hline 0.8 & 5000 & 0.4582 & 0.5224 & 0.7844 \tabularnewline & & (0.016)& (0.013)& (0.019)\tabularnewline \hline \end{tabular} \par\end{centering} \caption{Mean and standard deviation of estimates from 100 simulations for different $\rho$ and sample size. The standard (uncorrected) estimators (with previous tick and refresh time synchronization) and the corrected estimator, prescribed in this chapter, are reported.}\label{Table 1}
\end{table}

\begin{table}

\begin{tabular}{ccccc}
\hline
$\rho$ & $n$ & MSE (previous tick) & MSE (refresh time) & MSE (corrected)\tabularnewline
\hline
\hline
-0.4 & 800 & 0.0307 & 0.0195 & 0.0047\tabularnewline
\hline
0.2 & 800 & 0.0087 & 0.0078 & 0.0047\tabularnewline
\hline
0.8 & 800 & 0.112 & 0.0765 & 0.0027\tabularnewline
\hline
-0.4 & 2000 & 0.029 & 0.0179 & 0.0015\tabularnewline
\hline
0.2 & 2000 & 0.0085 & 0.0062 & 0.0021\tabularnewline
\hline
0.8 & 2000 & 0.1155 & 0.0755 & 0.0009\tabularnewline
\hline
-0.4 & 5000 & 0.0296 & 0.018 & 0.0006\tabularnewline
\hline
0.2 & 5000 & 0.007 & 0.004 & 0.0003\tabularnewline
\hline
0.8 & 5000 & 0.1117 & 0.077 & 0.0006\tabularnewline
\hline
\end{tabular}\caption{Calculated mean square error for the results in Table \ref{Table 1}.}\label{Table 2}

\end{table}

From the table, we see that both previous tick and refresh time sampling fail to capture the magnitude of true dependence. In fact the previous tick method is the worst choice for synchronization.

We carried out the same analysis with the $t$ copula, with different marginal distributions
with different degrees of freedom, which is a more realistic scenario for intraday financial data. The result is similar i.e. not only does
our prescribed correction give a good estimate but also the uncorrected method returns a biased estimate and the bias is significant. The result of 100 simulations with parameter -0.4 is summarized in Table \ref{Table 3}.

\begin{table}

\begin{tabular}{ccccc}
\hline
t copula (df 8) & mean  & sd  & mean  & sd \tabularnewline
marginals & uncorrected estimate & uncorrected estimate & corrected estimate & corrected estimate\tabularnewline
\hline
(t(5), t(7)) & -0.2623 & 0.036 & -0.3932 & 0.054\tabularnewline
(N(0,2), N(0,4)) & -0.264 & 0.038 & -0.3961 & 0.055\tabularnewline
(t(4), N(0,3)) & -0.2532 & 0.039 & -0.3801 & 0.059\tabularnewline
\hline
\end{tabular}\caption{Simulation for $t$ copula for different marginals with $\rho=-0.4$. The standard uncorrected estimates with refresh time sampling and the corrected estimates are reported.}\label{Table 3}
\end{table}

\subsection{Interval estimation of Kendall's Tau in non-Elliptical copula}\label{three methods subsection}
\begin{figure}
    \centering
    \includegraphics[width=0.5\textwidth]{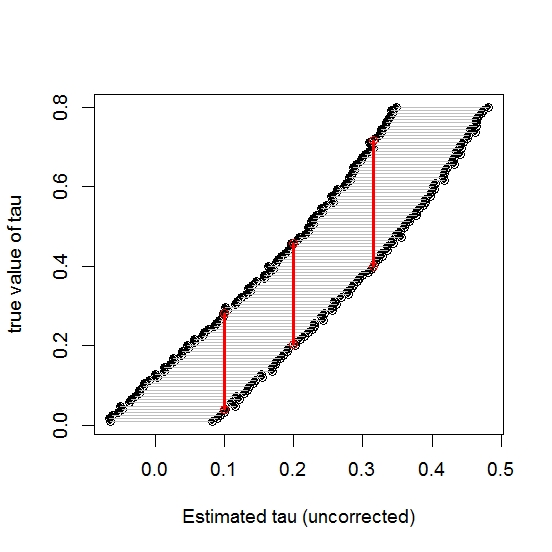}%
    \includegraphics[width=0.5\textwidth]{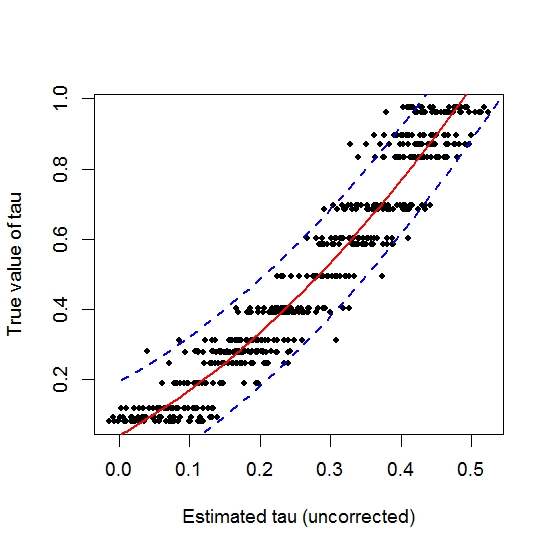}
    \caption{Left:Intervals in which estimated Kendall's tau lies in 95\% of times for Clayton copula, Right: Prediction interval for the regression line for Clayton copula.}
    \label{Clayton CI}
\end{figure}

: We take three approaches to interval estimation of the true Kendall's tau and applied those on simulated data from several Archimedean copulas. In the first approach, we follow the method described above and get the 95\% prediction interval. The blue dotted lines in the right panel of Figure. \ref{Clayton CI} show the prediction intervals for Clayton copula. \\  The second approach is similar to the first one, but we don't fit a regression line. Instead, for each true Kendall's tau, we plot the interval that contains the (under)estimated Kendall's tau 95\% of times.  In the left panel of Fig. \ref{Clayton CI}, we plot the intervals (horizontal) against the true Kendall's tau. Now we calculate the confidence interval for true Kendall's tau as the vertical interval corresponding to the estimated Kendall's tau (see the red vertical lines corresponding to 0.1, 0.2 and 0.32 in the figure).

In the third approach, we deliberately mis-specify the underlying copula as a Gaussian copula and use Theorem 1 to calculate the confidence interval (using the relation between correlation coefficient and Kendall's tau for elliptical copula, see eq. \ref{eq:elliptical_para_tau_relation}). The results of these three approaches to interval estimation are given in Table \ref{Table: 3 methods} in section \ref{sec:Simulation}.

The coverage probability and interval lengths from the three methods of interval estimation, described in section \ref{three methods subsection}, are shown in Table \ref{Table: 3 methods}. An important takeaway from this table comes from the last column which demonstrates that the effect of model misspecification can be quite serious. Note that the second method, being completely non-parametric, does not assume anything about the shape of the dependence function between the true parameter and the uncorrected estimate. The first method assumes a quadratic model. This assumption reduces computations by a huge amount. From the table we see that the coverage probability of the first method is always at least the target value of 95\%. So the assumption of a quadratic model does not compromise the coverage probability. The intervals are a little larger than the second method, so the first method is more conservative. Another observation is that the length of the intervals do not depend much on the value of the underlying parameter.

\begin{table}
\begin{center}
\begin{tabular}{ c c c c c }
 Copula & True Kendall's tau & 1st Method & 2nd Method & 3rd Method\\
    &    & (CP, IL) & (CP, IL) & (CP, IL)\\
    \hline
 Clayton & 0.1 & (.98,0.30) & (.95,0.19) & (.84,0.1)\\
 Clayton & 0.2 & (.98,0.31) & (.96,0.24) & (.77,0.3)\\
 Clayton & 0.3 & (.97,0.31) & (.97,0.25) & (.56,.29)\\
 Clayton & 0.5 & (.95,0.31) & (.97,0.29) & (.29,.26)\\
 Gumbel & 0.1 & (.98,0.30) & (.98,0.21) & (0.8,0.2)\\
 Gumbel & 0.2 & (.99,0.31) & (.94,0.24) & (.72,0.3)\\
 Gumbel & 0.3 & (.99,0.31) & (.95,0.25) & (.55,.29)\\
 Gumbel & 0.5 & (.95,0.31) & (.93,0.27) & (.23,.26)\\
 \hline
\end{tabular}
\caption{Coverage probability (CP) and Interval length (IL) for three methods.}\label{Table: 3 methods}
\end{center}
\end{table}

\section{Real data analysis }\label{sec:Data}

\begin{figure}
\begin{minipage}[t]{0.45\columnwidth}%
\includegraphics[scale=0.4]{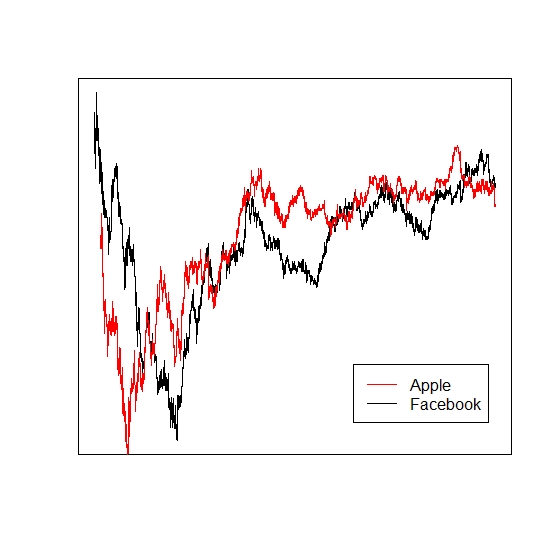}%
\end{minipage}\hfill{}%
\begin{minipage}[t]{0.45\columnwidth}%
\includegraphics[scale=0.4]{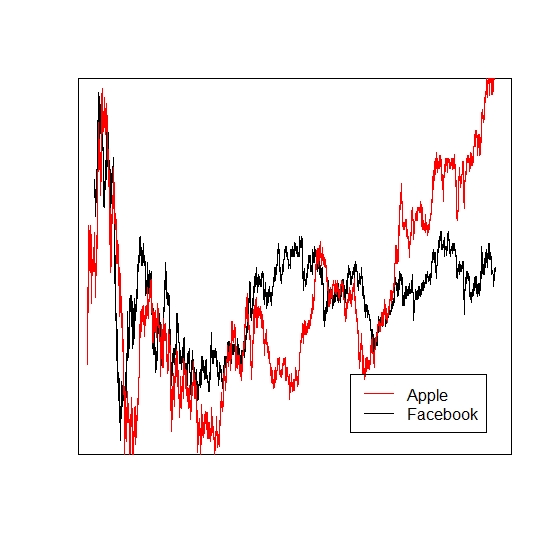}%
\end{minipage}\caption{Facebook (black) and Apple (gray) data after subtracting the mean for
two consecutive days 10.05.2017 and 11.05.2017}\label{Figure7}
\end{figure}

We analyze real financial intraday data to see which kind of copula is most likely to be encountered in practice. We use AIC to compare and select the best copula. In many of the cases, we find that the t-copula is a good choice to model bivariate intraday data. To see the impact of asynchronicity for real data we record the relative extent of correction to be undertaken. The intraday data for Apple and Facebook stocks are plotted in Figure \ref{Figure7}. These have been modelled by bivariate $t$ copula for three consecutive days. For all three days both the uncorrected and the corrected estimates are evaluated in Table \ref{Table4}. The percentage change in values of uncorrected and corrected estimates is reported in the third column. We notice that almost 30 to 35\% of data being lost or deleted after constructing the pairs by algorithm $\mathcal{A}_0$.

\begin{table}

\begin{tabular}{l|ccc|}
\cline{2-4}
 & uncorrected  & corrected & percentage change\tabularnewline
 & biased estimate & unbiased estimate & ($\frac{\mathrm{unbiased\ est.-biased\ est}}{\mathrm{biased\ est.}}\times100$)\tabularnewline
\cline{2-4}
 Day 1 & 0.098 & 0.141 & 43.87\%\tabularnewline
 Day 2 & 0.129 & 0.186 & 44.18\%\tabularnewline
 Day 3 & 0.111 & 0.159 & 43.24\%\tabularnewline
\cline{2-4}
\end{tabular}\caption{Copula estimation for the joint distribution of Apple and Facebook
data}\label{Table4}

\end{table}
We also perform the same analysis for a couple of other stocks and the results we obtained are very similar. For example, when we consider Amazon and Netflix on three nearly consecutive days, the percentage changes in copula parameter with $t$ copula are 41.75\%, 39.84\%, 42.76\% respectively.

\section{Conclusion and Future directions}\label{sec:Conclusions}
Both simulations and real data analysis clearly show that the impact of asynchronicity can be very serious if not tackled properly. We discuss some of the methods to circumvent the problem. Careful pre-processing of intraday data is necessary to model or infer about the underlying realities. We propose a consistent estimator of the correlation coefficient and more generally of elliptical copula function. For a more general class of copulas, where there is a one-one relation between the Kendall's tau and the copula, we suggest a way of estimating the copula parameter. Alongside the point estimates, three ways of interval estimation is discussed and compared. From the results it is evident that the impact of asynchronicity can be quite serious under model mis-specification. The real data analysis corroborates our findings. For the two chosen stocks, as the correlation is very less, the absolute change in the value after the correction is not much. But the relative change is significantly high, as we expected.

There are several directions in which one can extend this work. Firstly, we didn't assume the presence of microstructure noise. In the presence of noisy observations, the estimator may demand further modifications. The estimation procedure can be further challenging if the parameter is time-dependent. As time-dependent copula modelling is gaining popularity in financial data analysis, it is worthwhile to investigate the effect of asynchronicity in time-varying parameter estimation. Another question one can look into is that how asynchronicity affect the estimation of popular risk measures like Value at Risk (VaR).
\setcitestyle{numbers}
\bibliographystyle{agsm}
\bibliography{copulapaper}

\appendix
\section{Proofs}\label{Proofsec}
\subsection{Proof of Theorem \ref{thcoppara}:}
\begin{proof}
The conditional expectation: \[E[(X_{t(k_{i}^{1})}-X_{t(k_{i-1}^{1})})(Y_{t(k_{i}^{2})}-Y_{t(k_{i}^{2})})|(t(k_{i}^{1}),t(k_{i}^{2}),t(k_{i-1}^{1}),t(k_{i-1}^{2}))]\\ =I_iE(XY).\]
This is a consequence of the assumption $A_{2}$ (as illustrated in the examples in section \ref{sec:est}). \[E[(X_{t(k_{i}^{1})}-X_{t(k_{i-1}^{1})})^{2}|t(k_{i}^{1}),t(k_{i-1}^{1})] =(t(k_{i}^{1})-t(k_{i-1}^{1}))E(X^{2}).\]
Similarly, \[ E[(Y_{t(k_{i}^{2})}-Y_{t(k_{i-1}^{2})})^{2}|t(k_{i}^{2}),t(k_{i-1}^{2})]=(t(k_{i}^{2})-t(k_{i-1}^{2}))E(Y^{2}). \]
Then we have, $E(X_{t(k_{i}^{1})}-X_{t(k_{i-1}^{1})})^{2}=E(t(k_{i}^{1})-t(k_{i-1}^{1}))E(X^{2})$ and \\
$E(Y_{t(k_{i}^{2})}-Y_{t(k_{i-1}^{2})})^{2}=E(t(k_{i}^{2})-t(k_{i-1}^{2}))E(Y^{2})$.\\
Therefore, \begin{align*} \mathrm{Cor}((X_{t(k_{i}^{1})}-X_{t(k_{i-1}^{1})}),(Y_{t(k_{i}^{2})}-Y_{t(k_{i-1}^{2})})) & =\frac{E(XY)E(I_i)}{\sqrt{E(t(k_{i}^{1})-t(k_{i-1}^{1}))}E(X^{2})\sqrt{E(t(k_{i}^{2})-t(k_{i-1}^{2}))}E(Y^{2})}\\ & =\frac{\rho E(I_i)}{\sqrt{E(t(k_{i}^{1})-t(k_{i-1}^{1}))}\sqrt{E(t(k_{i}^{2})-t(k_{i-1}^{2}))}}. \end{align*}
Note that the estimate $\hat{\theta}$ is defined as
\[
\hat{\theta}=\sqrt{\frac{m_{1}m_{2}}{m(I)}}\hat{\rho},
\]
 where $\hat{\rho}=\hat{Cor}((X_{t(k_{i}^{1})}-X_{t(k_{i-1}^{1})}),(Y_{t(k_{i}^{2})}-Y_{t(k_{i-1}^{2})}))$.

Let us define the correction factor: $w=\frac{\sqrt{m_{1}m_{2}}}{m(I)}$.

Now,
\[
m_{1}\stackrel{P}{\rightarrow}E(t(k_{i}^{1})-t(k_{i-1}^{1}))
\]
 and
\[
m_{2}\stackrel{P}{\rightarrow}E(t(k_{i}^{1})-t(k_{i-1}^{1}))
\]
and
\[
m(I)\stackrel{P}{\rightarrow}E(I_{i}).
\]
Therefore, $w\stackrel{P}{\rightarrow}\gamma$, where
\[
\gamma=\frac{\sqrt{E(t(k_{i}^{1})-t(k_{i-1}^{1}))E(t(k_{i}^{2})-t(k_{i-1}^{2}))}}{E(I_{i})}.
\]
As $\hat{\rho}$ is sample correlation coefficient, we know that
\[
\sqrt{n}(\hat{\rho}-\rho)\stackrel{d}{\rightarrow}N(0,(1-\rho^{2})^{2}).
\]
By Slutsky's theorem we get,
\[
\sqrt{n}(w\hat{\rho}-\gamma\rho)\stackrel{d}{\rightarrow}N(0,\gamma^{2}(1-\rho^{2})^{2}).
\]
Using $\theta=\gamma\rho$ we then have,
\[
\sqrt{n}(\hat{\theta}-\theta)\stackrel{d}{\rightarrow}N(0,\gamma^{2}(1-\theta^{2}/\gamma^{2})^{2})=\gamma(1-\theta^{2}/\gamma^{2})N(0,1).
\]
Now we need to stabilize the variance.

Note that,
\[
\frac{1}{\gamma(1-\theta^{2}/\gamma^{2})}=\frac{1}{2\gamma}[\frac{1}{1-x/\gamma}+\frac{1}{1+x/\gamma}]
\]
Define, $f(x)=\frac{1}{2}[log(1+\frac{x}{\gamma})-log(1-\frac{x}{\gamma})].$
Then,
\[
f'(x)=\frac{1}{2\gamma}[\frac{1}{1+x/\gamma}+\frac{1}{1-x/\gamma}].
\]
Therefore a simple application of delta method implies that
\[
\sqrt{n}(f(\hat{\theta})-f(\theta))\stackrel{d}{\rightarrow}N(0,1).
\]
This completes the proof.
\end{proof}
\subsection{Proof of Theorem \ref{thcopfn}.}
\begin{proof}
 Note that $F(r^1,r^2)=C(F_{1}(r^1),F_{2}(r^2),\theta)$. \\So $\hat{C}=\hat{C}(\hat{F}_{1}(r^1_{t(k_{i}^{1})}),\hat{F}_{2}(r^2_{t(k_{i}^{1})}),\hat{\theta};\ i=1(1)n)$
is a consistent estimator for the copula $C$. But $R^2_{t(k_{i}^{1})}$'s
are unobserved, where $R^2_{t(k_{i}^{2})}$'s are actually observed.
Let us use the notation $\hat{F}_{2}(.)$ and $\tilde{F}_{2}(.)$
for the empirical distribution function of $R^2$ based on the observations
$\{r^2_{t(k_{i}^{1})}:i=1(1)n\}$ and $\{r^2_{t(k_{i}^{2})}:i=1(1)n\}$
respectively. Therefore to claim that the estimated copula based
on the paired data (observed) is consistent, we have to show that-\\
$|\hat{C}(\hat{F}_{1},\hat{F}_{2},\hat{\theta})-\hat{C}(\hat{F}_{1},\tilde{F}_{2},\hat{\theta})|\rightarrow0$
a.s.
Suppose
$\delta_{i}=|r^2_{t(k_{i}^{1})}-r^2_{t(k_{i}^{2})}|$. Note that by Assumption $\mathcal{A}_4$, $\delta = \mathrm{max}({\delta_{i}}) < M$.
Then,
\begin{align*}
|\tilde{F}_{2}(r)-\hat{F}_{2}(r)| & =|\frac{1}{n}\sum_{i=1}^{n}I(R^2_{t(k_{i}^{2})}\leq r)-\frac{1}{n}\sum_{i=1}^{n}I(R^2_{t(k_{i}^{1})}\leq r)|\\
  & \leq\frac{1}{n}\sum_{i=1}^{n}|I(R^2_{t(k_{i}^{2})}\leq r)-I(R^2_{t(k_{i}^{1})}\leq r)|\\
 & \leq\frac{1}{n}\sum_{i=1}^{n}I(R^2_{t(k^{2}_{i}})\in(r-\delta_{i},r+\delta_{i}))I(R^2_{t(k_{i}^{1})}\in(r-\delta_{i},r+\delta_{i}))\\
 & \leq\frac{1}{n}\sum_{i=1}^{n}I(R^2_{t(k_{i}^{2})}\in(r-\delta,r+\delta))I(R^2_{t(k_{i}^{1})}\in(r-\delta,r+\delta))
\end{align*}
This implies that
\begin{align*}
 & \sum_{n=1}^{\infty}P\big(|\tilde{F}_{2}(r)-\hat{F}_{2}(r)|>\eta\big)\\
 & \leq \sum_{n=1}^{\infty}P\Big(\frac{1}{n}\sum_{i=1}^{n}|I(R^2_{t(k_{i}^{1})}\in(r-\delta,r+\delta))I(R^2_{t(k_{i}^{2})}\in(r-\delta,r+\delta))|>\eta\Big)\\
 & \leq\sum_{n=1}^{\infty}\sum_{1\leq i, j\leq n}\frac{1}{n^{2}\eta^{2}}E(I(R^2_{t(k_{i}^{1})}\in(r-\delta,r+\delta))\times I(R^2_{t(k_{i}^{2})}\in(r-\delta,r+\delta))\times\\
 & \qquad \qquad \qquad I(R^2_{t(k_{j}^{1})}\in(r-\delta,r+\delta))\times I(R^2_{t(k_{j}^{2})}\in(r-\delta,r+\delta))\\
 & \leq\sum_{n=1}^{\infty}\frac{3}{\eta^{2} n^{2}}\sum_{i=1}^{n}E(I(R^2_{t(k_{i}^{1})}\in(r-\delta,r+\delta))\times I(R^2_{t(k_{i}^{2})}\in(r-\delta,r+\delta))) \\
 & =\sum_{n=1}^{\infty}\frac{3}{\eta^{2} n}P[(R^2_{t(k_{i}^{1})},R^2_{t(k_{i}^{2})})\in (r-\delta, r+\delta)\times (r-\delta, r+\delta)] \\
 & \leq \sum_{n=1}^{\infty}\frac{3}{\eta^{2} n}P[|R^2_{t(k^1_i)}-R^2_{t(k^2_i)}|\leq 2\delta]  \\
 & =\sum_{n=1}^{\infty}\frac{3}{\eta^{2}}O(\frac{1}{n^{1+\psi}}) \\
 & <\infty
\end{align*}
The second inequality is due to Chebyshev's inequality and the last equality is due to $\mathcal{A}_{4}$. The third inequality is a consequence of asynchronicity as for each $i$, there are at most two $j$'s (the preceding and the next) for which $(R^1_{t(k_{i}^{1})},R^2_{t(k_{i}^{2})})$ and $(R^1_{t(k_{j}^{1})},R^2_{t(k_{j}^{2})})$ are dependent.

Hence by Borel Cantelli Lemma,  $|\tilde{F}_{2}(r)-\hat{F}_{2}(r)|\stackrel{a.s}{\rightarrow0}$. Again as $|\hat{F}_{2}(r)-F_{2}(r)|\stackrel{a.s}{\rightarrow0}$ we have that,
$|\tilde{F}_{2}(r)-F_{2}(r)|\stackrel{a.s}{\rightarrow0}$.

Now we have to show that uniform convergence will hold in this case. That is, we want to show $\forall\epsilon>0,$ $\mathrm{sup}_r|\tilde{F}_{2}(r)-F(r)|<\epsilon$. For any given $\epsilon>0$ we have a finite partition of the real line $-\infty=z_{0}\leq z_{1}\leq z_{2}\leq\cdots\leq z_{k}=\infty$ such that $F(z_{i+1}^{-})-F(z_{i})\leq\epsilon$. This can be achieved
by taking $z_{0}=-\infty$ and $z_{j+1}=\mathrm{sup}\{z:\ F(z)\leq F(z_{j})+\epsilon\}$.
Then, $F(z_{j+1})\geq F(z_{j})+\epsilon$. Because if $F(z_{j+1})<F(z_{j})+\epsilon$
then by right continuity there exists a $\xi>0$ such that $F(z_{j+1}+\xi)<F(z_{j})+\epsilon$,
hence contradicting the definition of $z_{j+1}$. So between $z_{j}$ and
$z_{j+1}$, $F$ jumps at least $\epsilon$. This can happen at most
finite number of times, so $k<\infty$. By our definition of $z_{j+1}$,
we have $F(z_{j+1}-\xi)\leq F(z_{j})+\epsilon\ \forall \xi>0$. Hence
$F(z_{i+1}^{-})-F(z_{i})\leq\epsilon$. \\ If $y\in[z_{i},z_{i+1})$
then we have,
\begin{align*}
\tilde{F}(z_{i}) & \leq\tilde{F}(r)\leq\tilde{F}(z_{i+1}^{-})\quad \mathrm{and}\\
F(z_{i}) & \leq F(r)\leq F(z_{i+1}^{-})\\
\implies & \tilde{F}(z_{i})-F(z_{i+1}^{-})\leq\tilde{F}(r)-F(r)\leq\tilde{F}(z_{i+1}^{-})-F(z_{i})\\
\implies & \tilde{F}(z_{i})-F(z_{i})+F(z_{i})-F(z_{i+1}^{-})\leq\tilde{F}(r)-F(r)\\
 & \leq\tilde{F}(z_{i+1}^{-})-F(z_{i+1}^{-})+F(z_{i+1}^{-})-F(z_{i})\quad \forall n>n_{i}\\
\implies-2\epsilon & \leq\tilde{F}(z_{i})-F(z_{i})+F(z_{i})-F(z_{i+1}^{-})\leq\tilde{F}(r)-F(r)\\
 & \leq\tilde{F}(z_{i+1}^{-})-F(z_{i+1}^{-})+F(z_{i+1}^{-})-F(z_{i})\\
 & \leq2\epsilon  \quad \forall n>n_{i}\\
\implies-2\epsilon & \leq\tilde{F}(r)-F(r)\leq2\epsilon \quad \forall r\in\mathbb{R}, \quad\forall n>\mathrm{max}(n_{i})\\
\implies & |\tilde{F}(r)-F(r)|\leq2\epsilon
\end{align*}
 Now using properties of copula we can clearly see that
\begin{align*}
 & |C(F_{1}(r^1),F_{2}(r^2))-C(\hat{F}_{1}(r^1),\tilde{F}_{2}(r^2))|\\
 & \leq|C(F_{1}(r^1),F_{2}(r^2))-C(\hat{F}_{1}(r^1),F_{2}(r^2))|+|C(\hat{F}_{1}(r^1),F_{2}(r^2))-C(\hat{F}_{1}(r^1),\tilde{F}_{2}(r^2))|\\
 & \leq|\hat{F}_{1}(r^1)-F_{1}(r^1)|+|\tilde{F}_{2}(r^2)-F_{2}(r^2)|
\end{align*}

As both $\hat{F}_{1}$ and $\tilde{F}_{2}$ are uniformly convergent
to $F_{1}$ and $F_{2}$ respectively, the result follows.
\end{proof}

\subsection{Proof of Theorem \ref{thdataloss}}

\begin{proof}

\textit{(a)} If you fix the point $t(k_{i}^{1})$
then there can be two situations depending on the position of $t(k_{i}^{2})$
as illustrated in Figure \ref{avgoverlap}.

\begin{figure}
\includegraphics[scale=0.5]{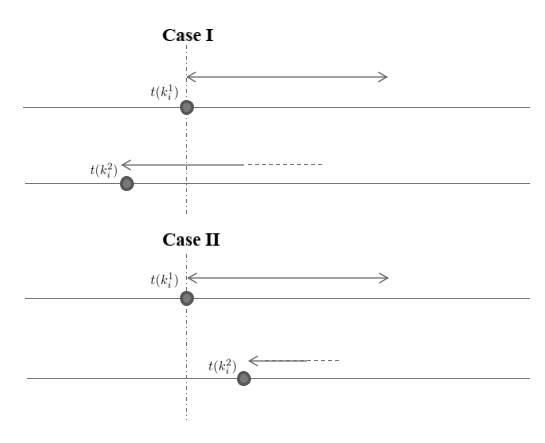}\caption{ Two situations depending on the position of $t(k_{i}^{2})$}\label{avgoverlap}
\end{figure}
Firstly consider case 1. We are interested in the overlapping interval. As
illustrated in Figure \ref{avgol1}, define $S_{1}$ as the first interarrival
of the first stock after $t(k_{i}^{1})$, $T'_{1}$ as the first
interarrival of the 2nd stock after $t(k_{i}^{2})$ and $T_{1}$ as the first interarrival of the 2nd stock after $t(k_{i}^{1})$ i.e.
if we start observing the process only from $t(k_{i}^{1})$ then the
first arrival time for the second stock. As the arrival processes
are Poisson processes, distributions of $T_{1}$ and $T'_{1}$ are
same (due to memory-less property). Denote all the subsequent interarrivals
as $S_{i}:i=2,3,...$ and $T_{i}:i=2,3,...$.

\begin{figure}
\includegraphics[scale=0.3]{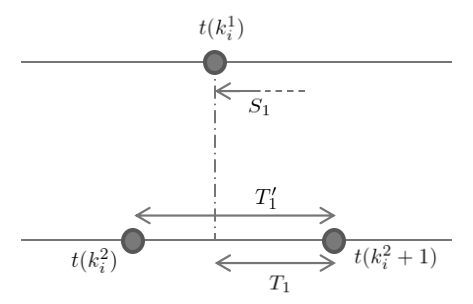}\caption{An illustration of the first case.}\label{avgol1}
\end{figure}
Now (for case 1),
\begin{align*}
I= \begin{cases} & \sum_{i=1}^{k}T_{i}  \quad\mathrm{if} \quad\mathrm{for}\ k(\geq1)\ \mathrm{s.t.}\ \sum_{i=1}^{k}T_{i}<S_{1}<\sum_{i=1}^{k+1}T_{i}\\
 & \sum_{i=1}^{k}S_{i}\quad\mathrm{if}\quad\mathrm{for\ }k(\geq1)\ \mathrm{s.t.}\ \sum_{i=1}^{k}S_{i}<T_{1}<\sum_{i=1}^{k+1}S_{i}
 \end{cases}
\end{align*}
Therefore (for case 1),
\[
E(I)=\sum_{n=1}^{\infty}\Bigg[E\Big(\sum_{i=1}^{n}T_{i}\Big)P\Big(\sum_{i=1}^{n}T_{i}<S_{1}<\sum_{i=1}^{n+1}T_{i}\Big)+E\Big(\sum_{i=1}^{n}S_{i}\Big)P\Big(\sum_{i=1}^{n}S_{i}<T_{1}<\sum_{i=1}^{n+1}S_{i}\Big)\Bigg]
\]
Now, $E\big(\sum_{i=1}^{n}T_{i}\big)=nE(T_{1})=\frac{n}{\lambda_{2}}$ and $E\big(\sum_{i=1}^{n}S_{i}\big)=nE(S_{1})=\frac{n}{\lambda_{1}}$. \\
Define $p_{n}=P\big(\sum_{i=1}^{n}T_{i}<S_{1}<\sum_{i=1}^{n+1}T_{i})$
and $q_{n}=P\big(\sum_{i=1}^{n}S_{i}<T_{1}<\sum_{i=1}^{n+1}S_{i}\big)$.
So,
\[
E(I)=\sum_{n=1}^{\infty}\Big[\frac{n}{\lambda_{1}}p_{n}+\frac{n}{\lambda_{2}}q_{n}\Big]
\]
\begin{align*}
q_{n} & =P\Big(\sum_{i=1}^{n}S_{i}<T_{1}<\sum_{i=1}^{n+1}S_{i}\Big)\\
 & =P\Big(T_{1}<\sum_{i=1}^{n+1}S_{i})-P(T_{1}<\sum_{i=1}^{n}S_{i}\Big)
\end{align*}
Note that $T_{1}\sim exp(\lambda_{2})\equiv Gamma(1,\lambda_{2})$
and $\sum_{i=1}^{n}S_{i}\sim Gamma(n,\lambda_{1})$.
Therefore,
\begin{align*}
P\Big(\sum_{i=1}^{n}S_{i} & >T_{1}\Big)=P\Bigg(\frac{\sum_{i=1}^{n}S_{i}}{\lambda_{1}}>\frac{(\lambda_{1}+\lambda_{2})T_{1}}{\lambda_{1}\lambda_{2}}-\frac{T_{1}}{\lambda_{2}}\Bigg)\\
 & =P\Bigg(\frac{\sum_{i=1}^{n}S_{i}}{\lambda_{1}}+\frac{T_{1}}{\lambda_{2}}>\frac{(\lambda_{1}+\lambda_{2})T_{1}}{\lambda_{1}\lambda_{2}}\Bigg)\\
 & =P\Bigg(\frac{T}{\frac{\sum_{i=1}^{n}S_{i}}{\lambda_{1}}+\frac{T_{1}}{\lambda_{2}}}<\frac{\lambda_{1}\lambda_{2}}{\lambda_{1}+\lambda_{2}}\Bigg)\\
 & =P\Bigg(\frac{T/\lambda_{2}}{\frac{\sum_{i=1}^{n}S_{i}}{\lambda_{1}}+\frac{T_{1}}{\lambda_{2}}}<\frac{\lambda_{1}}{\lambda_{1}+\lambda_{2}}\Bigg)\\
 & =P\Big(Z_{n}<\frac{\lambda_{1}}{\lambda_{1}+\lambda_{2}}\Big)
\end{align*}
where $Z_{n}=\frac{T/\lambda_{2}}{\frac{\sum_{i=1}^{n}S_{i}}{\lambda_{1}}+\frac{T_{1}}{\lambda_{2}}}\sim Beta(1,n)$
Thereffore,
\[
q_{n}=F_{Z_{n+1}}\Big(\frac{\lambda_{1}}{\lambda_{1}+\lambda_{2}}\Big)-F_{Z_{n}}\Big(\frac{\lambda_{1}}{\lambda_{1}+\lambda_{2}}\Big)
\]
Similarly,
\[
p_{n}=F_{Z_{n+1}}\Big(\frac{\lambda_{2}}{\lambda_{1}+\lambda_{2}}\Big)-F_{Z_{n}}\Big(\frac{\lambda_{2}}{\lambda_{1}+\lambda_{2}}\Big)
\]
Similarly we can derive for case 2 by interchanging the role of $X$ and $Y$. See Figure \ref{intarrvl}. Hence for case 2 we have,
\[
E(I)=\sum_{n=1}^{\infty}\Big[\frac{n}{\lambda_{1}}p_{n}+\frac{n}{\lambda_{1}}q_{n}\Big]
\]
Now two cases are equally likely, $P(\mathrm{Case}\ 1)=\frac{1}{2}=P(\mathrm{Case\ 2})$.
Therefore combining both the cases we have, $E(I)=\frac{1}{2}\sum_{n=1}^{\infty}n[(\frac{1}{\lambda_{1}}+\frac{1}{\lambda_{2}})(p_{n}+q_{n})]$.\\
\begin{figure}
\includegraphics[scale=0.5]{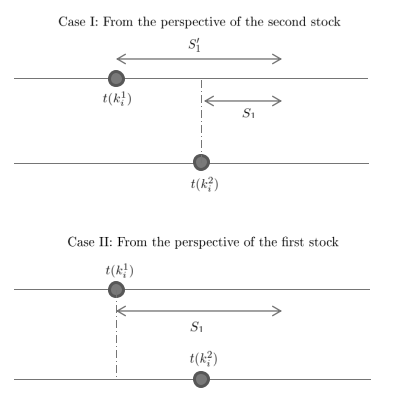}\caption{ A case with two perspectives.}\label{intarrvl}
\end{figure}
\textit{(b)} Now we have to calculate $E(t(k_{i}^{1})-t(k_{i-1}^{1}))$.\\
For case 1:
Define,
\begin{align*}
N= \begin{cases} & 1\quad\mathrm{if}\quad\mathrm{for}\ k(\geq1)\ \mathrm{s.t.}\  \sum_{i=1}^{k}T_{i}<S_{1}<\sum_{i=1}^{k+1}T_{i}\\
 & k\quad\mathrm{if}\quad\mathrm{for}\ k(\geq1)\ \mathrm{s.t.}\ \sum_{i=1}^{k}S_{i}<T_{1}<\sum_{i=1}^{k+1}S_{i}
 \end{cases}
\end{align*}
Then, $E(t(k_{i}^{1})-t(k_{i-1}^{1}))=\frac{1}{\lambda_{1}}\eta_{1}$, where $E(N)=\eta_{1}$
\begin{align*}
E(N) & =\sum_{k=1}^{\infty}\Bigg[P\Big(\sum_{i=1}^{k}T_{i}<S_{1}<\sum_{i=1}^{k+1}T_{i}\Big)+kP\Big(\sum_{i=1}^{k}S_{i}<T_{1}<\sum_{i=1}^{k+1}S_{i}\Big)\Bigg]\\
 & =\sum_{k=1}^{\infty}\Bigg[\Big\{P\Big(S_{1}<\sum_{i=1}^{k+1}T_{i}\Big)-P\Big(S_{1}<\sum_{i=1}^{k}T_{i}\Big)\Big\}+k\Big\{P\Big(T_{1}<\sum_{i=1}^{k+1}S_{i}\Big)-P\Big(T_{1}<\sum_{i=1}^{k}S_{i}\Big)\Big\}\Bigg]\\
 & =\sum_{k=1}^{\infty}\Bigg[\Big\{F_{B(1,k+1)}\Big(\frac{\lambda_{2}}{\lambda_{1}+\lambda_{2}}\Big)-F_{B(1,k)}\Big(\frac{\lambda_{2}}{\lambda_{1}+\lambda_{2}}\Big)\Big\}+k\Big\{F_{B(1,k+1)}\Big(\frac{\lambda_{1}}{\lambda_{1}+\lambda_{2}}\Big)-F_{B(1,k)}\Big(\frac{\lambda_{1}}{\lambda_{1}+\lambda_{2}}\Big)\Big\}\Bigg]
\end{align*}\\
For case 2 the derivation is similar.\\
\textit{(c)} Following similar derivation as part \textit{(b)} $E(t(k_{i}^{2})-t(k_{i-1}^{2}))=\frac{1}{\lambda_{2}}\eta_{2}$,
where $\eta_{2}$ is the mean of $N'$ defined below,
\begin{align*}
N'=\begin{cases} & 1\quad\mathrm{if}\quad\mathrm{for}\ k(\geq1)\ \mathrm{s.t.} \ \sum_{i=1}^{k}S_{i}<T_{1}<\sum_{i=1}^{k+1}S_{i}\\
 & k\quad\mathrm{if}\quad\mathrm{for}\ k(\geq1) \ \mathrm{s.t.} \ \sum_{i=1}^{k}T_{i}<S_{1}<\sum_{i=1}^{k+1}T_{i}
 \end{cases}
\end{align*}
 A similar derivation will establish-
\[
\eta_{2}=\sum_{k=1}^{\infty}\Bigg[\Big\{F_{B(1,k+1)}\Big(\frac{\lambda_{1}}{\lambda_{1}+\lambda_{2}}\Big)-F_{B(1,k)}\Big(\frac{\lambda_{1}}{\lambda_{1}+\lambda_{2}}\Big)\Big\}+k\Big\{F_{B(1,k+1)}\Big(\frac{\lambda_{2}}{\lambda_{1}+\lambda_{2}}\Big)-F_{B(1,k)}\Big(\frac{\lambda_{2}}{\lambda_{1}+\lambda_{2}}\Big)\Big\}\Bigg]
\]
\end{proof}

\subsection{Proof of Lemma \ref{lemma1}}
\begin{proof}
\begin{align*} E(\mathrm{sign}(A)) & =E(\mathrm{sign}(A)| \mathrm{sign}(A)\neq \mathrm{sign}(B))P(\mathrm{sign}(A)\neq \mathrm{sign}(B)) \\ & +E(\mathrm{sign}(A)| \mathrm{sign}(A)= \mathrm{sign}(B))P(\mathrm{sign}(A)=\mathrm{sign}(B)) \\ & =E(\mathrm{sign}(A)| \mathrm{sign}(A)\neq \mathrm{sign}(B))\frac{1}{2}+E(\mathrm{sign}(A)| \mathrm{sign}(A)= \mathrm{sign}(B))\frac{1}{2} \end{align*}
As $Y(\eta_{1})+Y(\epsilon_{1})-Y(\eta_{2})-Y(\epsilon_{2})$ is independent
of $Y(u_{1})-Y(u_{2})$ and symmetric around $0$, clearly \begin{align*} & E(\mathrm{sign}(A)|\mathrm{sign}(A)\neq \mathrm{sign}(B)) \\ & =E(\mathrm{sign}(A)|\mathrm{sign}(Y(u_{1})-Y(u_{2})\neq \mathrm{sign}(Y(\eta_{1})+Y(\epsilon_{1})-Y(\eta_{2})-Y(\epsilon_{2}))) \\ & =E(\mathrm{sign}(A)|\mathrm{sign}(Y(u_{1})-Y(u_{2})=\mathrm{sign}(Y(\eta_{1})+Y(\epsilon_{1})-Y(\eta_{2})-Y(\epsilon_{2})))\\ & = E(\mathrm{sign}(A)|\mathrm{sign}(A)= \mathrm{sign}(B)) \end{align*}
So, $E(\mathrm{sign}(A)|\mathrm{sign}(A)\neq\mathrm{sign}(B))=E(\mathrm{sign}(A))$.
\end{proof}

\subsection{Proof of Theorem \ref{1stth}}
\begin{proof}

Due to symmetry of two configurations, it is enough to prove the theorem
for one configuration. Let us consider the
4th configuration. Here $I_{1}=u_{1}$, $I_{2}=u_{2}$, $I_{1}^{c}=\epsilon_{1}+\eta_{1}$
and $I_{2}^{c}=\epsilon_{2}+\eta_{2}$.

The Kendall's tau for this nonsynchronous configuration, as defined
in section 1, is: \begin{align*} \rho_{\tau} & =E(\mathrm{sign}(X_{1}-X_{2})(Y_{1}-Y_{2}))\\ & =E(\mathrm{sign}(X_{1}(u_{1})-X_{2}(u_{2}))(Y_{1}(\epsilon_{1})+Y_{1}(u_{1})+Y_{1}(\eta_{1})-Y_{2}(\epsilon_{2})-Y_{2}(u_{2})-Y_{2}(\eta_{2})))\\ & =E(\mathrm{sign}\{(X_{1}(u_{1})-X_{2}(u_{2}))(Y_{1}(u_{1})-Y_{2}(u_{2}))\\ & \ \ +(X_{1}(u_{1})-X_{2}(u_{2}))(Y_{1}(\epsilon_{1})+Y_{1}(\eta_{1})-Y_{2}(\epsilon_{2})-Y_{2}(\eta_{2}))\}) \end{align*}

According to our notation, $A=(X_{1}(u_{1})-X_{2}(u_{2}))(Y_{1}(u_{1})-Y_{2}(u_{2}))$
and $B=(X_{1}(u_{1})-X_{2}(u_{2}))(Y_{1}(\epsilon_{1})+Y_{1}(\eta_{1})-Y_{2}(\epsilon_{2})-Y_{2}(\eta_{2}))$.
So, $\rho_{\tau}=E(\mathrm{sign}(A+B))$ and $\tilde{\rho_{\tau}}=E(\mathrm{sign}(A))$.
 Let us
denote the region, where $\mathrm{sign}(A)\neq\mathrm{sign}(A+B)$, by $N$ i.e. $N=\{\mathrm{sign}(A)\neq\mathrm{sign}(B)\ \&|B|>|A|\}$.

\begin{align*} E(\mathrm{sign}(A+B)) &= E(\mathrm{sign}(A+B)|N^{c})P(N^{c})+E(\mathrm{sign}(A+B)|N)P(N)\\ & = E(\mathrm{sign}(A)|N^{c})P(N^{c})-E(\mathrm{sign}(A)|N)P(N) \end{align*}
But \begin{align*} E(\mathrm{sign}(A)) & = E(\mathrm{sign}(A)|N^{c})P(N^{c})+E(\mathrm{sign}(A)|N)P(N) \end{align*}
By Lemma \ref{lemma2}, $|E(\mathrm{sign}(A)|N)|>0$. This implies $|E(\mathrm{sign}(A+B))|<|E(\mathrm{sign}(A))|$.
\end{proof}

\subsection{Proof of Theorem \ref{oldth}}
\begin{proof}
Due to symmetry of two configurations, it is enough to prove the theorem
for one configuration. We consider the case of Figure $\ref{Figure4}$
(4th configuration). Here $I_{1}=u_{1}$, $I_{2}=u_{2}$, $I_{1}^{c}=\epsilon_{1}+\eta_{1}$
and $I_{2}^{c}=\epsilon_{2}+\eta_{2}$.

The Kendall's tau for this nonsynchronous configuration, as defined
in section 1, is: \begin{align*} \rho_{\tau} & =E(\mathrm{sign}(X_{1}-X_{2})(Y_{1}-Y_{2}))\\ & =E(\mathrm{sign}(X_{1}(u_{1})-X_{2}(u_{2}))(Y_{1}(\epsilon_{1})+Y_{1}(u_{1})+Y_{1}(\eta_{1})-Y_{2}(\epsilon_{2})-Y_{2}(u_{2})-Y_{2}(\eta_{2})))\\ & =E(\mathrm{sign}\{(X_{1}(u_{1})-X_{2}(u_{2}))(Y_{1}(u_{1})-Y_{2}(u_{2}))\\ & \ \ +(X_{1}(u_{1})-X_{2}(u_{2}))(Y_{1}(\epsilon_{1})+Y_{1}(\eta_{1})-Y_{2}(\epsilon_{2})-Y_{2}(\eta_{2}))\}) \end{align*}

According to our notation, $A=(X_{1}(u_{1})-X_{2}(u_{2}))(Y_{1}(u_{1})-Y_{2}(u_{2}))$
and $B=(X_{1}(u_{1})-X_{2}(u_{2}))(Y_{1}(\epsilon_{1})+Y_{1}(\eta_{1})-Y_{2}(\epsilon_{2})-Y_{2}(\eta_{2}))$.
So,

\begin{align*} \rho_{\tau} & =E(\mathrm{sign}(A+B))\\ & =E(\mathrm{sign}(A)|\mathrm{sign}(A+B)=\mathrm{sign}(A))P(\mathrm{sign}(A+B)=\mathrm{sign}(A))\\ & -E(\mathrm{sign}(A)|\mathrm{sign}(A+B)\neq \mathrm{sign}(A))P(\mathrm{sign}(A+B)\neq\mathrm{sign}(A))\\ & =E(\mathrm{sign}(A)|\mathrm{sign}(A+B)={sign}(A))p-E(\mathrm{sign}(A)|\mathrm{sign}(A+B)\neq \mathrm{sign}(A))(1-p)\\ & =p(E(\mathrm{sign}(A)|\mathrm{sign}(A+B)=\mathrm{sign}(A))+E(\mathrm{sign}(A)|\mathrm{sign}(A+B)\neq \mathrm{sign}(A)))\\ & -E(\mathrm{sign}(A)|\mathrm{sign}(A+B)\neq \mathrm{sign}(A)) \end{align*}

and \begin{align*} \tilde{\rho}_{\tau} & =E(\mathrm{sign}(A)|\mathrm{sign}(A+B)=\mathrm{sign}(A))P(\mathrm{sign}(A+B)=\mathrm{sign}(A))\\ & +E(\mathrm{sign}(A)|\mathrm{sign}(A+B)\neq \mathrm{sign}(A))P(\mathrm{sign}(A+B)\neq\mathrm{sign}(A))\\ & =p(E(\mathrm{sign}(A)|\mathrm{sign}(A+B)=\mathrm{sign}(A))-E(\mathrm{sign}(A)|\mathrm{sign}(A+B)\neq \mathrm{sign}(A)))\\ & +E(\mathrm{sign}(A)|\mathrm{sign}(A+B)\neq \mathrm{sign}(A)) \end{align*}

where $\tilde{\rho}_{\tau}$ is the true Kendall's tau and $p=P(\mathrm{sign}(A+B)=\mathrm{sign}(A))$.

Therefore, \begin{equation} \rho_{\tau}+\tilde{\rho}_{\tau}=2p(E(\mathrm{sign}(A)|\mathrm{sign}(A+B)=\mathrm{sign}(A)))\label{eq:1-1} \end{equation}

Now we have to calculate the probability $p$.

\begin{align*} p & =P(\mathrm{sign}(A+B)=\mathrm{sign}(A))\\ & =P(\{\mathrm{sign}(A)=\mathrm{sign}(B)\}\ or\ \{\mathrm{sign}(A)\neq\mathrm{sign}(B)\ \&\ |A|>|B|\})\\ & =P(\{\mathrm{sign}(A)=\mathrm{sign}(B)\})+P(\{\mathrm{sign}(A)\neq\mathrm{sign}(B)\ \&\ |A|>|B|\})\\ & =P(\{\mathrm{sign}(A)=\mathrm{sign}(B)\})+P(\{\mathrm{sign}(A)\neq\mathrm{sign}(B)\})P(|A|>|B|)\\ & =\frac{1}{2}+\frac{1}{2}P(|A|>|B|) \end{align*}

The 3rd step of the above derivation is justified because $\{\mathrm{sign}(A)=\mathrm{sign}(B)\}$
and $\{\mathrm{sign}(A)\neq\mathrm{sign}(B)\ \&\ |A|>|B|\}$ are clearly
independent. Also independence of $\{\mathrm{sign}(A)\neq\mathrm{sign}(B)$
and $\{|A|>|B|\}$ is self evident which results in step 4. Note that
$\mathrm{sign}(A)$ and $\mathrm{sign}(B)$ will depend on $\mathrm{sign}(Y_{1}(u_{1})-Y_{2}(u_{2}))$
and $\mathrm{sign}(Y_{1}(\epsilon_{1})+Y_{1}(\eta_{1})-Y_{2}(\epsilon_{2})-Y_{2}(\eta_{2}))$.
Due to independent increment property, $\mathrm{sign}(Y_{1}(u_{1})-Y_{2}(u_{2}))$
and $\mathrm{sign}(Y_{1}(\epsilon_{1})+Y_{1}(\eta_{1})-Y_{2}(\epsilon_{2})-Y_{2}(\eta_{2}))$
are independent. Therefore $P(\mathrm{sign}(A)=\mathrm{sign}(B))=P(\mathrm{sign}(A)\neq\mathrm{sign}(B))=\frac{1}{2}$.

Let us denote the events $C=\{\mathrm{sign}(A)=\mathrm{sign}(B)\}$
and $D=\{\mathrm{sign}(A)\neq\mathrm{sign}(B)\&|A|>|B|\}$. Note that
$\tilde{\rho}_{\tau}=E(\mathrm{sign}(A))=E(\mathrm{sign}(A)|C)$ as
condition on the event $C$ does not influence the expected value
of $\mathrm{sign}(A)$.

\begin{align*} E(\mathrm{sign}(A)|\mathrm{sign}(A+B) & =\mathrm{sign}(A))\\ & =\frac{E(\mathrm{sign}(A)|C)P(C)+E(\mathrm{sign}(A)|D)P(D)}{P(C\cup D)}\\ & =\frac{\tilde{\rho}_{\tau}/2+E(\mathrm{sign}(A)|D)(P(|A|>|B|)/2)}{p} \end{align*}

Inserting this to \ref{eq:1-1} we get \begin{align*} \rho_{\tau}+\tilde{\rho}_{\tau} & =2p\frac{\tilde{\rho}_{\tau}/2+E(\mathrm{sign}(A)|D)(P(|A|>|B|)/2)}{p}\\ \implies\rho_{\tau} & =E(\mathrm{sign}(A)|D)P(|A|>|B|) \end{align*}

Hence we proved the result.
\end{proof}

\section{Positive and negative connection}

Suppose $X$ and $Y$ are positively associated. And $(X_{1},Y_{1})$
and $(X_{2},Y_{2})$ are two identical pairs. Then bigger the value
of $Y_{1}$ from $Y_{2}$, bigger the probability of $\{X_{1}>X_{2}\}$
would expected to be. This good expectation is formalized in the following
definition.

\begin{defn}
\begin{doublespace}
$X$ is said to be \textit{positively connected} to $Y$ if $\forall M>0$,
\begin{align*}
P(U>0|V>M) & \geq P(U>0|0<V<M)\\
 & \geq P(U>0|-M<V<0)\\
 & \geq P(U>0|V<-M)
\end{align*}

$X$ is said to be \textit{negatively connected} to $Y$ if $\forall M>0$,
the signs of the above inequalities are reversed.
\end{doublespace}
\end{defn}

\begin{doublespace}
Note that, the definition is not symmetric for $X$ and $Y$. $X$
is positively connected to $Y$ does not mean that $Y$ is positively
connected to $X$.
\end{doublespace}

\textbf{Definition:}
If $X$ is \textit{positively (or negatively) connected} to $Y$ and
$Y$ is \textit{positively (or negatively) connected} to $X$, then
we call- there is a \textit{positive (or negative) connection} between
$X$ and $Y$.

\begin{doublespace}
It is easy to see that if $X$ and $Y$ have \textit{positive (/negative)
connection}, then \textit{assumption 1}, \textit{assumption 2} and
\textit{assumption 2'} are satisfied. This is because $X_{1}-X_{2}|Y_{1}-Y_{2}\stackrel{d}{=}X_{2}-X_{1}|Y_{2}-Y_{1}$.

Therefore this stronger but reasonable assumption unifies the previous
assumptions and enable us to state the following theorem.
\end{doublespace}

\begin{thm}
\begin{doublespace}
\label{unifthm} Under the assumption that returns of two stocks have
positive (/negative) connection, for the pairs with 1st and 4th configuration,
\[ |\tilde{\rho_{\tau}}|>|\rho_{\tau}|) \] and  \[ \mathrm{sign}(\tilde{\rho_{\tau}})=\mathrm{sign}(\rho_{\tau}) \]
where $\rho_{\tau}$ is the Kendall's tau calculated on the paired
data with 1st and 4th configurations, i.e. \textup{$\rho_{\tau}=E(\mathrm{sign}(X_{1}-X_{2})(Y_{1}-Y_{2}))$,}
where $(X_{1},Y_{1})$ and $(X_{2},Y_{2})$ are independent pairs
of the same configurations.\textup{ }
\end{doublespace}
\end{thm}
\end{document}